\documentclass[%
 reprint,
 amsmath,amssymb,
 aps,
]{revtex4-2}

\usepackage{graphicx}

\usepackage{xcolor}

\newcommand{\rf}[1]{(\ref{#1})}

\newcommand{\beq}{\begin{equation}}
\newcommand{\eeq}{\end{equation}}
\newcommand{\beqr}{\begin{eqnarray}}
\newcommand{\eeqr}{\end{eqnarray}}

\newcommand{\lb}[1]{\label{#1}}

\newcommand{\bc}{\begin{center}}
\newcommand{\ec}{\end{center}}
\newcommand{\ct}[1]{\cite{#1}}

\begin{document}

\title{Perturbation approach in Heisenberg equations for lasers}

\author{Igor E. Protsenko}
\email{procenkoie@lebedev.ru}

\author{Alexander V. Uskov}

\affiliation{%
Quantum Electronic division, Lebedev Physical Institute,
Moscow 119991, Russia
}%

\date{\today}

\begin{abstract}
Nonlinear Heisenberg-Langevin equations  are solved analytically by operator Fourier-expansion for the laser in the light emitting diode (LED)  regime. Fluctuations of populations of lasing levels are taken into account as perturbations. Spectra of operator products are  calculated as convolutions, preserving Bose commutations for the lasing field operators. It is found that fluctuations of population  significantly affect spontaneous and stimulated emissions into the lasing mode, increase the radiation rate, the number of lasing photons and broad the spectrum of a bad cavity thresholdless and the superradiant lasers. The method can be applied to various resonant systems in quantum optics. 
\begin{description}
\item[Keywords] Heisenberg equations, superradiant lasers
\end{description}
\end{abstract}

\maketitle

\section{\label{Sec1}Introduction}
Operator Heisenberg-Langevin equations (HLE), as quantum Maxwell-Bloch  equations, are widely used in quantum optics and laser physics   \ct{PhysRevA.51.R3426}. They are applied for modelling devices and processes in nonlinear optic \ct{PhysRevA.49.2065, DEMETER20131203}, lasers \ct{Scully,PhysRevA.47.1431,RevModPhys.68.127,PhysRevA.59.1667}, generation of non-classical light \ct{FABRE1992215}, qbits \ct{PhysRevA.93.063820} and  other quantum phenomena  \ct{Kundu2019} making an impotrant part of physics \ct{Ackemann2005}.  
HLEs are in the background of various theoretical methods of quantum optics as   the input-output theory \ct{PhysRevA.46.2766,PhysRevA.30.1386} and the cluster expansion method \ct{Jahnke,PhysRevA.75.013803}. 

HLE for lasers and resonant optical systems are often nonlinear in operators, which makes difficult to solve them analytically. This paper continues and extends the research of \ct{Protsenko_2021}  on analytical solving HLEs for lasers.     

Several methods of solving HLE are proposed  \ct{PhysRevA.96.062108, Ara_jo_2019,LANGOUCHE1980301,Langner2015,Castej_n_2003,Mista_2001,ARIMITSU1991163}. Relatively simple and widespread  method of solving HLE in  quantum optics and laser physics \ct{Scully, 1071986,1071726,PhysRevA.47.1431,PhysRevA.59.1667,RevModPhys.68.127} is a  generalization of the  perturbation approach of the classical oscillation theory \ct{Osc_th}. This is the linearization of HLE around mean values of operators and solving linear equations for  operators of small perturbations. 

Consider, for example, the nonlinear term $\hat{a}\hat{N}_e$ in Eq.~\rf{MBE_2} of the laser model in  Section~\ref{Sec2}, where $\hat{a}$ is a Bose-operator of the lasing field amplitude and $\hat{N}_e$ is the operator of the population of excited states of lasing transitions. $\hat{N}_e$ can be separated on the mean $N_e$ and fluctuations $\delta\hat{N}_e$: $\hat{N}_e = N_e+\delta\hat{N}_e$. Supposing that the contribution of fluctuations $\delta\hat{N}_e$ is small and can be neglected, we approximately replace   $\hat{a}\hat{N}_e$ by the term $\hat{a}{N}_e$ linear in the operator $\hat{a}$. Then the stationary HLE for the laser in Section~\ref{Sec2} are linearized and can be solved as in \ct{PhysRevA.59.1667,Andre:19,Protsenko_2021}  at a weak excitation of the laser, when the laser does not generate coherent radiation, and the mean amplitude of the lasing field $a=0$. This approach reproduces well-known results, as the laser linewidth \ct{Protsenko_2021} and leads to new results, as the collective Rabi splitting \ct{Andre:19}, but it must be extended for considering population fluctuations at the weak excitation of the laser.

Similar way the laser HLE can be linearized and solved for a high excitation, when the laser does generate coherent radiation, so $\hat{a} \approx a +\delta\hat{a}$, where  $\delta\hat{a}$ is the operator of a small perturbation  \ct{PhysRevA.47.1431,RevModPhys.68.127,Protsenko_2021}. In this case population fluctuations are taken into account and lead to well-known relaxation oscillations peaks in the intensity fluctuation spectra \ct{1071726}  and to the prediction of such peaks in the field spectra of the bad cavity nanolasers  \ct{Protsenko_2021}.

Direct generalization of the standard  perturbation approach 
for considering population fluctuations at a low excitation 
meets difficulties. Consider, for example, the laser at a weak excitation, when the mean laser field  $a=0$. Following the standard procedure of the classical perturbation theory we neglect $\delta\hat{N}_e$ and find a zero-order solution $\hat{a} = \hat{a}_0$ \ct{PhysRevA.59.1667,Andre:19,Protsenko_2021}. Next  
we must replace  $\hat{a}\delta\hat{N}_e$ with the linear term $\hat{a}_0\delta\hat{N}_e$ and obtain linear equations with the time dependent  operator coefficients, like $\hat{a}_0$. It is unclear  how to solve such equations.

To overcome such a difficulty,  in \ct{Protsenko_2021} we  replace   $\hat{a}_0$ in $\hat{a}_0\delta\hat{N}_e$ by $\sqrt{n}$, where $n$ is the mean photon number. This approach 
makes a "smooth transition" between the high and the low excitation of the laser, but  remained without a justification  for the low excitation in \ct{Protsenko_2021}. It was mentioned in \ct{Protsenko_2021},  that the approach is good, if the population fluctuations with the low excitation are negligibly small (we will see, that this is not always the case).  Features  of lasing, found in \ct{Protsenko_2021} due to the population fluctuations at the low excitation, need a prove with more rigorous approach.    

One purpose of this work is to extend the analysis of \ct{Protsenko_2021} and consider population fluctuations rigorously at the low excitation, when the laser works in the LED regime. We will correct some results of \ct{Protsenko_2021} related  with population fluctuations  in the LED regime.  

We  outlined above, that it is difficult to take into account population fluctuations in the nonlinear laser HLEs  at the low excitation with the standard perturbation approach.  Another purpose of the paper is to formulate a perturbation approach for  solving nonlinear  stationary HLEs  at the low excitation of the laser  in the first order on population fluctuations. 

Only a few methods can be applied in the higher order on quantum perturbations as, for example, a cluster expansion method \ct{Jahnke,PhysRevA.75.013803}. It lets to find mean values of high-order correlations of products of operators, but it does not calculate spectra of optical fields. 
Path integral formalism can be used in some problems of nonlinear and quantum optics \ct{PhysRevA.26.451,doi:10.1063/5.0055815}. However, it is applied mostly to systems with quadratic Hamiltonian, i.e. to linear systems. 
Quantum perturbation theory in time is often applied for analysis of non-stationary processes in nonlinear optics \ct{Andrews:20}, and it is restricted by short periods of time, when the effect of nonlinear terms is negligibly small.

Here we consider  the population fluctuations as a perturbation using the operator Fourier-expansion, and express power spectra of the operator products  as convolutions of spectra of multipliers in the product.  

An important part of the method is  preserving   commutation relations for  Bose operators of the field. This lets us to take into account quantum fluctuations in the field with a small number of photons.

Because of the dissipation and fluctuations, oscillation spectra of resonant systems are bands centered at mode frequencies. We suppose, as usual,  that the width of the band is much smaller than the mode frequency and use a rotating wave approximation (RWA)  \ct{Fujii2017}. 

As usual, we suppose that the laser interacts with incoherent "white noise" baths of broad spectra.

We demonstrate the method on the example of  quantum model of  single mode laser with homogeneously broaden active medium of two-level emitters, the same as in \ct{Andre:19, Protsenko_2021}. We suppose a large number of emitters $N_0\gg1$ and consider the LED radiation regime at a weak excitation of the laser, when the mean number $n$ of lasing photons is small $n<1$ or of the order of 1, so the laser does not generate coherent radiation.  
 
We will show that population fluctuations increase, at certain conditions, the radiation rate into the lasing mode; increase the number of lasing photons and broad  lasing spectra. This can be seen, most clearly, in lasers with low quality cavities and large gain, where population fluctuations are high and collective effects, as a superradiance, are important \ct{Khanin, Belyanin_1998,Koch_2017}. Such superradiant lasers  have been experimentally realized, for example,  with cold alkaline earth atoms~\ct{PhysRevX.6.011025, PhysRevA.96.013847,PhysRevA.81.033847,PhysRevA.98.063837},  rubidium atoms~\ct{Bohnet}, and with quantum dots~\ct{Jahnke}.
 
Quantum models of a laser have been presented in many papers and books as, for example, \ct{Lax_book1966,SLW,Scully}. Among popular methods of the laser theory are the linearization of Heisenberg-Langevin equations around the steady state \ct{Lax_book1966,PhysRevA.47.1431,RevModPhys.68.127}, solving the master equation for density matrix \ct{Scully} or Lindblad master equation \ct{doi:10.1063/1.5115323}. The method proposed here has not been used before.   

Usual perturbation theory with the linearization of operator equations on small fluctuations around the steady states  is widely used in the laser quantum rate equation theory \ct{Coldren,Henry1986,1071726,doi:10.1063/1.5022958,McKinstrie:20}. Quantum rate equations for lasers are valid with the adiabatic elimination of the polarization of the lasing media. The method, presented here, does not require the adiabatic elimination of polarization, so it can be applied for the modelling of lasers with bad cavities and collective effects. 

In this paper we do not  provide  rigorous mathematical justification of the method, in particular, we do not prove its conversion to the exact solution. Our aim is to demonstrate  basic physical ideas  and to show the   application of the method.  
We will use general properties of Heisenberg representation and well-known results of quantum mechanics \ct{Landau_Quant} for the derivation of the mathematical part of the method in Appendixes~\ref{App_op} and~\ref{AppB}.   

We demonstrate the method on the example of the laser model  described in Section~\ref{Sec2}. There we derive the laser HLE and obtain from them equations for  Fourier-component operators. 

In Section \ref{Sec_zero_order} we  apply the  perturbation approach to the laser model  in the zero-order approximation, when population fluctuations are neglected. 

In Section \ref{Sec_first_order} we solve the laser equations, taking into account population fluctuations in the first-order approximation. We demonstrate the important parts of the method:  calculation of the spectrum of the operator product with convolutions and  preserving Bose-commutation relations for the lasing field operator. 

Section \ref{Sec5} presents and discuses results related with the effect of population fluctuations on the lasing in the LED regime at low excitation. We show that population fluctuations increase the spontaneous and the stimulated emission rates into the lasing mode leading to the increase of the number of lasing photons, they broad the lasing field spectra, but do not lead to narrow peaks in the field spectra found in \ct{Protsenko_2021}. Such peaks are  the consequence of the application of the standard perturbation approach at the low excitation.

Results are summarized in Conclusion. 

Appendix~\ref{App_op} shows the Fourier-expansion for operators, Appendix~\ref{AppB} calculates  the spectrum of the operator product, Appendix~\ref{DC_app} calculates diffusion coefficients. Appendix~\ref{Sigma_eqs} presents equations for population fluctuations  for calculation of the population fluctuation spectrum and the justification of the approximation \rf{pop_fl_eq}. 
\section{\label{Sec2} equations for two-level laser}
\begin{figure}[thb]\bc
\includegraphics[width=7cm]{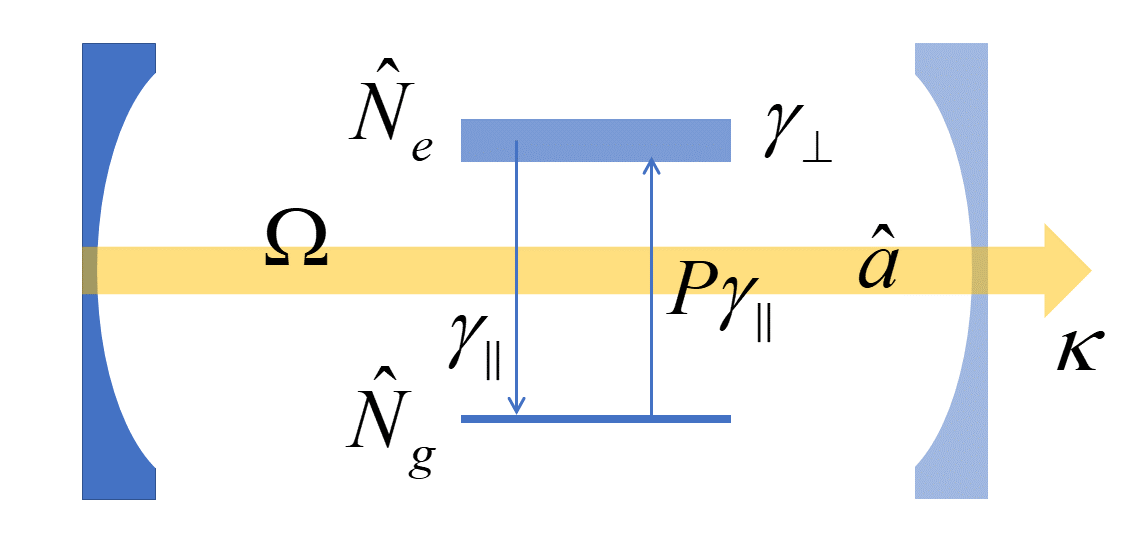}
\caption{Scheme of the two-level laser. Upper levels of emitters with the  population operator $\hat{N}_e$, decay to the low levels with the rate $\gamma_{\parallel}$   and pumped with the rate $\gamma_{\parallel}P$  from the low levels with the population $\hat{N}_g$. The width of the lasing transition is $\gamma_{\perp}$. Lasing mode described by Bose-operator  $\hat{a}$ decays through the semitransparent mirror with the rate  $\kappa$ and resonantly interacts with lasing transitions of two-level emitters with the vacuum Rabi frequency $\Omega$.   }
\label{Fig1}\ec
\end{figure}

We consider a quantum model of a single mode homogeneously broaden laser in the stationary regime with $N_0\gg 1$ two-level identical emitters, the same as in \ct{Andre:19,Protsenko_2021},  shown schematically in Fig.~\ref{Fig1}. 
%
%
Lasing transitions  are in the exact resonance with the cavity mode with the optical frequency $\omega_0$.  $\hat{a}(t){{e}^{-i{{\omega }_{0}}t}}$   is Bose-operator of the lasing mode, the operator $\hat{a}(t)$  of complex amplitude is changed much slowly than ${{e}^{-i{{\omega }_{0}}t}}$.  

Hamiltonian of the laser, written in the interaction picture with the carrier frequency  $\omega_0$ and in the RWA approximation, is
\beq
        H=i\hbar \Omega \sum\limits_{i=1}^{{{N}_{0}}}{{{f}_{i}}\left( {{{\hat{a}}}^{+}}{{{\hat{\sigma }}}_{i}}-\hat{\sigma }_{i}^{+}\hat{a} \right)}+\hat{\Gamma }. \lb{las_H}
\eeq
Here $\Omega$  is the vacuum Rabi frequency, $f_i$  describes the difference in couplings of different emitters with the lasing mode. 
$\hat{\sigma}_i$  is a lowing operator of i-th emitter, $\hat{\Gamma}$  describes the interaction of the mode and emitters with the white noise baths of the environment. 

Commutation relations for operators are 
\[
        \left[ \hat{a},{{{\hat{a}}}^{+}} \right]=1, \hspace{0.25cm} \left[ {{{\hat{\sigma }}}_{i}},\hat{\sigma }_{j}^{+} \right]=\left( \hat{n}_{i}^{g}-\hat{n}_{i}^{e} \right){{\delta }_{ij}},\]\beq \left[ {{{\hat{\sigma }}}_{i}},\hat{n}_{j}^{e} \right]=\left[ \hat{n}_{j}^{g},{{{\hat{\sigma }}}_{i}} \right]={{\delta }_{ij}}{{\hat{\sigma }}_{i}},\lb{cr_H}
\eeq
where $\hat{n}_{j}^{e}$  and $\hat{n}_{j}^{g}$  are operators of populations of the upper and the low levels of i-th emitter, $\delta_{ij}$ is Kronecker symbol. 

We introduce operators  $\hat{v}$ and $\hat{N}_{e,g}$  of the polarization and  populations of all emitters
\beq
\hat{\nu }=\sum\limits_{i=1}^{{{N}_{0}}}{{{f}_{i}}{{{\hat{\sigma }}}_{i}}} \hspace{0.5cm} {{\hat{N}}_{e,g}}=\sum\limits_{i=1}^{{{N}_{0}}}{\hat{n}_{i}^{e,g}}. \lb{v_and_Ne}
\eeq
Using commutation relations \rf{cr_H} and Hamiltonian \rf{las_H}  we write Maxwell-Bloch equations for $\hat{a}$, $\hat{v}$ and ${\hat{N}}_{e}$ 
\begin{subequations}\lb{MBE_0}\beqr
  \dot{\hat{a}}&=&-\kappa \hat{a}+\Omega \hat{v}+{{{\hat{F}}}_{a}} \lb{MBE_1}\\
 \dot{\hat{v}}&=&-\left( {{\gamma }_{\bot }}/2 \right)\hat{v}+\Omega f\hat{a}\left( 2{{{\hat{N}}}_{e}}-{{N}_{0}} \right)+{{{\hat{F}}}_{v}} \lb{MBE_2}\\ 
 {{{\dot{\hat{N}}}}_{e}}&=&-\Omega \hat{\Sigma }+{{\gamma }_{\parallel }}\left[ P\left( {{N}_{0}}-{{{\hat{N}}}_{e}} \right)-{{{\hat{N}}}_{e}} \right]+{{{\hat{F}}}_{{{N}_{e}}}}, \lb{MBE_3} 
\eeqr\end{subequations}
where
\beq
     \hat{\Sigma }={{\hat{a}}^{+}}\hat{v}+{{\hat{v}}^{+}}\hat{a},   \lb{sigma_def}
\eeq
$\kappa$, ${\gamma }_{\bot }$ and ${\gamma }_{\parallel }$  are decay rates,  $P{\gamma }_{\parallel }$ is the pump rate, ${{\hat{F}}_{\alpha }}$ with the index $\alpha =\left\{ a,v,{{N}_{e}} \right\}$  are Langevin forces. Total number of emitters is preserved, so ${\hat{N}}_{e}+{\hat{N}}_{g}={N}_{0}$.

In Eqs.~\rf{MBE_0} and below we approximate $f_{i}^{2}\approx f=N_{0}^{-1}\sum\limits_{i=1}^{{{N}_{0}}}{f_{i}^{2}}$ and use notations with a “hat” for operators and without a hat for mean values as, for example, ${{N}_{e}}=\left\langle {{{\hat{N}}}_{e}} \right\rangle$.

We separate mean values and fluctuations in population operators ${{\hat{N}}_{e,g}}={{N}_{e,g}}+\delta {{\hat{N}}_{e,g}}$, in $\hat{\Sigma} = \Sigma + \delta\hat{\Sigma}$, insert them into  Eqs.\rf{MBE_0} and write
\begin{subequations}\lb{MBE_St}\beqr
  \dot{\hat{a}}&=&-\kappa \hat{a}+\Omega \hat{v}+{{{\hat{F}}}_{a}} \lb{MBE_St1} \\ 
 \dot{\hat{v}}&=&-\left( {{\gamma }_{\bot }}/2 \right)\hat{v}+\Omega f\left( \hat{a}N+2\hat{a}\delta {{{\hat{N}}}_{e}} \right)+{{{\hat{F}}}_{v}}. \lb{MBE_St2}\\
 {{{\delta\dot{\hat{N}}}}_{e}}&=&-\Omega \delta\hat{\Sigma }-\gamma_P\delta\hat{N}_e+\hat{F}_{N_e},  \lb{MBE_St3}
\eeqr\end{subequations}
where $\gamma_P=\gamma_{\parallel }(P+1)$. With the derivation of Eqs.~\rf{MBE_St3} we take
\beq
0=-\Omega \Sigma +{{\gamma }_{\parallel }}\left[ P\left( {{N}_{0}}-{{N}_{e}} \right)-{{N}_{e}} \right].\lb{st_sigma}
\eeq
In Eq.~\rf{MBE_St2} and below $N={{N}_{e}}-{{N}_{g}}$ is the mean population inversion. 

We take the stationary mean photon number $n = \left<\hat{a}^+\hat{a}\right>$ and find from Eq.~\rf{MBE_St1}
\beq
0=-2\kappa n +   \Omega\Sigma. \lb{eql_0}
\eeq
Eq.~\rf{eql_0} and Eq.~\rf{st_sigma} lead to the energy conservation law
\beq
        2\kappa n=  \gamma_{\parallel}[P(N_0-N_e)-N_e]. \lb{eql_1}
\eeq

In the next sections we consider population fluctuations $\delta\hat{N}_e$ as a perturbation and solve the stationary Eqs.~\rf{MBE_St} approximately  using Fourier-expansion for operators  
\beq
\hat{\alpha}(t)=\frac{1}{\sqrt{2\pi }}\int_{-\infty }^{\infty }{\hat{\alpha}(\omega )}e^{-i\omega t}d\omega, \lb{Fu_exp_gn}
\eeq
where  $\hat{\alpha}$ denotes an operator $\hat{\alpha} = \hat{a},\hat{v},...$.  In particular,  $\hat{\alpha}$ can be the product of operators $\hat{a}\delta {{{\hat{N}}}_{e}}$. $\hat{\alpha}(\omega )$ is Fourier-component of the operator $\hat{\alpha}(t)$. $\hat{\alpha}(\omega )$ can be expressed through $\hat{\alpha}(t)$ by the reverse Fourier-transform, see more about the operator Fourier-expansion in Appendix~\ref{App_op}. 

In the stationary case
\beq
\left<\hat{\alpha}^+(\omega)\hat{\alpha}(\omega')\right> = S_{\alpha^+\alpha}(\omega)\delta(\omega+\omega'), \lb{spectrum}
\eeq
where $S_{\alpha^+\alpha}(\omega)$ is a power spectrum of fluctuations, corresponding to $\hat{\alpha}(t)$. We will find power spectra solving equations for Fourier-component operators and using relations as Eq.~\rf{spectrum}. Similar way of calculations of field spectra can be found in the literature, for example, in \ct{1072058,Henry1986,PhysRevB.97.125406}.  It can be shown, that  $S_{\alpha^+\alpha}(\omega)$ in Eq.~\rf{spectrum} is a Fourier-component of the auto-correlation function $\left<\hat{\alpha}^+(t+\tau)\hat{\alpha}(t)\right>$ in accordance with Wiener–Khinchin theorem \ct{wiener1964time,champeney_1987}.

Fourier-expansion for operators is widely used in laser physics and quantum optics \ct{PhysRevA.47.1431,RevModPhys.68.127,PhysRevA.59.1667,PhysRevA.30.1386,1072058,Henry1986,doi:10.1063/1.5022958} as well as in the classical stochastic theory \ct{Rytov1987}. However, the Fourier-expansion of a stochastic function  is not well-defined \ct{wiener1964time,champeney_1987}, so quite often the  calculation of  power spectra, as $S_{\alpha^+\alpha}(\omega)$, is carried out without the use of Fourier-component operators. Instead, one calculates a time-dependent autocorrelation function and then applies the Wiener–Khinchin theorem     \ct{1071986,PhysRevA.94.053813,Kirton_2018,Maier:14}. In our opinion,  the calculation of spectra in the stationary case with Fourier-component operators and the formula \rf{spectrum} (see examples in   \ct{1072058,1071726,doi:10.1063/1.5022958}) is more easy, than with the Wiener–Khinchin theorem. However  the operator Fourier-expansion \rf{Fu_exp_gn} must be justified, so in Appendix~\ref{App_op} we make the operator Fourier-expansion \rf{Fu_exp_gn} basing on quantum-mechanical relations in Heisenberg picture in the stationary case.

Making Fourier-expansion \rf{Fu_exp_gn} in Eqs.~\rf{MBE_St}  we obtain algebraic equations for Fourier-component operators 
\begin{subequations}\lb{FC_0}\beqr
0&=&\left( i\omega -\kappa  \right)\hat{a}(\omega )+\Omega \hat{v}(\omega )+{{{\hat{F}}}_{a}}(\omega ) \lb{FC_1}\\ 
 0&=&\left( i\omega -{{\gamma }_{\bot }}/2 \right)\hat{v}(\omega )+ \lb{FC_2}\\ & & \Omega f\left[ \hat{a}(\omega )N+2{{\left( \hat{a}\delta {{{\hat{N}}}_{e}} \right)}_{\omega }} \right]+{{{\hat{F}}}_{v}}(\omega ).\nonumber\\
0&=&(i\omega-\gamma_P)\delta\hat{N}_e(\omega)-\Omega \delta\hat{\Sigma }(\omega)+\hat{F}_{N_e}(\omega).  \lb{FC_3} 
\eeqr\end{subequations}
Here ${{\left( \hat{a}\delta {{{\hat{N}}}_{e}} \right)}_{\omega }}$ is a Fourier-component of the operator product $\hat{a}(t)\delta\hat{N}_e(t)$. 

Correlations for Fourier-components of Langevin forces ${{\hat{F}}_{\alpha }}(\omega )$, ${{\hat{F}}_{\beta }}(\omega )$ are 
\beq
\left\langle {{{\hat{F}}}_{\alpha }}(\omega ){{{\hat{F}}}_{\beta }}(\omega ') \right\rangle =2{{D}_{\alpha \beta }}\delta (\omega +\omega '), \lb{diff_coeff}
\eeq
where $2{{D}_{\alpha \beta }}$ is a spectral power density of the bath noise or a diffusion coefficient. Diffusion coefficients  
\beq
2D_{a{{a}^{+}}}^{{}}=2\kappa, \hspace{0.5cm}      2D_{{{a}^{+}}a}=0	\lb{dc_h_osc}
\eeq
correspond to the lasing mode - harmonic oscillator \ct{PhysRevA.46.2766,PhysRevA.30.1386}, they remain the same in any order of our approach. We choose diffusion coefficients $2D_{{{v}^{+}}v}^{(i)}$ and $2D_{v{{v}^{+}}}^{(i)}$ such, that Bose commutation relations for the operator $\hat{a}$ of the lasing mode will be preserved in $i=0,1...$ order of the approximation on population fluctuations. 
\section{\label{Sec_zero_order} Zero-order approximation}
In the zero-order approximation we neglect population fluctuations \ct{PhysRevA.59.1667,Andre:19,Protsenko_2021}. We drop the term ${{\left( \hat{a}\delta {{{\hat{N}}}_{e}} \right)}_{\omega }}$ in Eq.~\rf{FC_2},  and take Langevin force ${{\hat{F}}_{v}}(\omega )=\hat{F}_{v}^{(0)}(\omega )$ with diffusion coefficients
\beq
	2D_{{{v}^{+}}v}^{(0)}=f{{\gamma }_{\bot }}{{N}_{e}}, \hspace{0.5cm}        2D_{v{{v}^{+}}}^{(0)}=f{{\gamma }_{\bot }}{{N}_{g}}. \lb{dc_1_app}
\eeq
These diffusion coefficients are found at the absence of population fluctuations in Appendix~\ref{DC_app}. 

In the zero-order approximation  $\hat{a}={{\hat{a}}_{0}}$.  
We solve the set of Eqs~\rf{FC_1}, \rf{FC_2}, taken without ${{\left( \hat{a}\delta {{{\hat{N}}}_{e}} \right)}_{\omega }}$, and find 
\beq
{{\hat{a}}_{0}}(\omega )=\frac{\left( {{\gamma }_{\bot }}/2-i\omega  \right){{{\hat{F}}}_{a}}(\omega )+\Omega {{{\hat{F}}}_{v}}(\omega )}{s(\omega)},\lb{a_00}
\eeq
where 
\beq
s(\omega)=\left( i\omega -\kappa  \right)\left( i\omega -{{\gamma }_{\bot }}/2 \right)-(\kappa\gamma_{\perp}/2)N/N_{th}, \lb{s_small}
\eeq
and ${{N}_{th}}=\kappa {{\gamma }_{\bot }}/2{{\Omega }^{2}}f$ is a threshold population inversion found in the semiclassical laser theory \ct{Andre:19,Protsenko_2021}. 

The spectrum ${{n}_{0}}(\omega )$ of the lasing field satisfies 
\beq\left\langle \hat{a}_{0}^{+}(\omega ){{{\hat{a}}}_{0}}(\omega ') \right\rangle ={{n}_{0}}(\omega )\delta \left( \omega +\omega ' \right).\lb{sp_n0_def}
\eeq
We calculate ${{n}_{0}}(\omega )$ from Eqs.~\rf{a_00}, \rf{sp_n0_def}  and using  diffusion coefficients \rf{dc_h_osc}, \rf{dc_1_app}
\beq
{{n}_{0}}(\omega )=\frac{(\kappa \gamma _{\bot }^{2}/2){{N}_{e}}/{{N}_{th}}}{S(\omega )},\lb{n0_sp}
\eeq
where $S(\omega ) = |s(\omega )|^2$. The mean photon number ${{n}_{0}}={{(2\pi )}^{-1}}\int\limits_{-\infty }^{\infty }{{{n}_{0}}(\omega )}d\omega $ is
\beq{{n}_{0}}=\frac{{{N}_{e}}}{(1+2\kappa /{{\gamma }_{\bot }})({{N}_{th}}-N)}.\lb{mean_n1}
\eeq
To ensure that Bose-commutation relations $\left\langle \left[ {{{\hat{a}}}_{0}},\hat{a}_{0}^{+} \right] \right\rangle =1$ are satisfied, we find the spectrum  ${{\left( {{n}_{0}}+1 \right)}_{\omega }}$ such that  $\left\langle {{{\hat{a}}}_{0}}(\omega )\hat{a}_{0}^{+}(\omega ') \right\rangle ={{\left( {{n}_{0}}+1 \right)}_{\omega }}\delta \left( \omega +\omega ' \right)$
\beq
	{{\left( {{n}_{0}}+1 \right)}_{\omega }}=\frac{2\kappa ({{\omega }^{2}}+\gamma _{\bot }^{2}/4)+(\kappa \gamma _{\bot }^{2}/2){{N}_{g}}/{{N}_{th}}}{S(\omega )},	\lb{sp_nand1}
\eeq
and the spectrum of the commutator $\left\langle \left[ {{{\hat{a}}}_{0}}(\omega ),\hat{a}_{0}^{+}(\omega ') \right] \right\rangle ={{\left[ {{{\hat{a}}}_{0}},\hat{a}_{0}^{+} \right]}_{\omega }}\delta \left( \omega +\omega ' \right)$ 
\beq
	{{\left[ {{{\hat{a}}}_{0}},\hat{a}_{0}^{+} \right]}_{\omega }}= {{\left( {{n}_{0}}+1 \right)}_{\omega }} - n_0(\omega).	\lb{sp_com}
\eeq
Calculation shows that ${{(2\pi )}^{-1}}\int\limits_{-\infty }^{\infty }{{{\left[ {{{\hat{a}}}_{0}},\hat{a}_{0}^{+} \right]}_{\omega }}}d\omega =1$, so Bose commutation relations for $\hat{a}_0$ are satisfied. 
\section{\label{Sec_first_order} First-order approximation}
In the first-order approximation we denote $\hat{a}={{\hat{a}}_{1}}$, keep in Eq.~\rf{FC_2} the term ${{\left( \hat{a}_0\delta {{{\hat{N}}}_{e}} \right)}_{\omega }}$ with $\hat{a}$ replaced by $\hat{a}_0$  and take Langevin force ${{\hat{F}}_{v}}(\omega )=\hat{F}_{v}^{(1)}(\omega )$ with diffusion coefficients
\beqr
2D_{{{v}^{+}}v}^{(1)}&=&f{{\gamma }_{\bot }}\left[ {{N}_{e}}+{{N}_{1}}(\omega ) \right] \lb{d_coeff_1}\\ 2D_{{}v{v}^{+}}^{(1)}&=&f{{\gamma }_{\bot }}\left[ {{N}_{g}}-{{N}_{1}}(\omega ) \right]. \nonumber
\eeqr
${{N}_{1}}(\omega )$ in Eqs.~\rf{d_coeff_1} is added for satisfing Bose commutation relations  $\left<[\hat{a}_1,\hat{a}_1^+]\right> = 1$.    
Expressions~\rf{d_coeff_1} are written such, that the sum  $2D_{{{v}^{+}}v}^{(1)}+2D_{v{{v}^{+}}}^{(1)}$ does not depend on ${{N}_{1}}(\omega )$ and, therefore, on population fluctuations, as it is shown in Appendix~\ref{DC_app}. This is why the same $N_1$ appears in both diffusion coefficients $2D_{v^+v}^{(1)}$ and $2D_{vv^+}^{(1)}$.

Solving the set of Eqs.~\rf{FC_1}  and \rf{FC_2} with $\left(\hat{a}_0\delta \hat{N}_e\right)_{\omega}$ and $\hat{F}_{v}^{(1)}(\omega )$ instead of $\left(\hat{a}\delta \hat{N}_e\right)_{\omega}$ and ${{\hat{F}}_{v}}(\omega )$, respectively, we find the  Fourier-component operator
\beq
{{\hat{a}}_{1}}(\omega )= \hat{a}_0^{(1)}+\frac{\kappa\gamma_{\bot }}{N_{th}}\frac{{{\left( {{{\hat{a}}}_{0}}\delta {{{\hat{N}}}_{e}} \right)}_{\omega }}}{s(\omega )}. \lb{a_1} 
\eeq
where $\hat{a}_0^{(1)}(\omega)$ is  given by Eq.~\rf{a_00} with ${{\hat{F}}_{v}}(\omega )=\hat{F}_{v}^{(1)}(\omega )$.

Now we find  ${{\left( {{{\hat{a}}}_{0}}\delta {{{\hat{N}}}_{e}} \right)}_{\omega }}$ and ${{N}_{1}}(\omega )$. We consider the spectrum ${{S}_{{{a}_{0}}{{N}_{e}}}}(\omega )$ of the operator product ${{\hat{a}}_{0}}\delta {{\hat{N}}_{e}}$ 
\beq
	\left\langle {{\left( {{{\hat{a}}}_{0}^+}\delta {{{\hat{N}}}_{e}} \right)}_{\omega }}{{\left( {{{\hat{a}}}_{0}}\delta {{{\hat{N}}}_{e}} \right)}_{\omega '}} \right\rangle ={{S}_{{{a}_{0}}{{N}_{e}}}}(\omega )\delta \left( \omega +\omega ' \right). 	\lb{sp_op_prod}
\eeq
We calculate ${{S}_{{{a}_{0}}{{N}_{e}}}}(\omega )$ neglecting  cumulants in correlations, as in a well-known cumulant-neglect closure method in the classical statistical theory \ct{Wu_1984,10.1115/1.3173083} and in the quantum  cluster-expansion method \ct{PhysRevA.75.013803}. In these methods the mean of, for example, four-operator products is approximated by the sum of products of the non-zero  two-operator means.  In case of Eq.~\rf{sp_op_prod} this  is
\beqr
&\left< \hat{a}_0^+(\omega_1)\delta \hat{N}_e(\omega_2)  \hat{a}_0(\omega_3)\delta \hat{N}_e(\omega_4)  \right>&\nonumber\\ \approx&\left< \hat{a}_0^+(\omega_1)\hat{a}_0(\omega_3)\right>\left<\delta \hat{N}_e(\omega_2)  \delta \hat{N}_e(\omega_4)  \right>,& \lb{cumulant_no}
\eeqr
since $\left< \hat{a}_0^+(\omega_1)\delta \hat{N}_e(\omega_2)\right> =0$ and $\left< \hat{a}_0(\omega_1)\delta \hat{N}_e(\omega_2)\right> =0$ at the low excitation of the laser.

It is shown in Appendix~\ref{AppB}, that ${{S}_{{{a}_{0}}{{N}_{e}}}}(\omega )$ calculated with the approximation \rf{cumulant_no} is a convolution ${{S}_{{{a}_{0}}{{N}_{e}}}}(\omega )={{\left( {{n}_{0}}*{{\delta }^{2}}{{N}_{e}} \right)}_{\omega }}$,
\beq
	{{\left( {{n}_{0}}*{{\delta }^{2}}{{N}_{e}} \right)}_{\omega }}= \frac{1}{2\pi }\int\limits_{-\infty }^{\infty }{{{n}_{0}}(\omega -\omega ')}{{\delta }^{2}}{{N}_{e}}(\omega ')d\omega ', 	\lb{conv}
\eeq
where ${{\delta }^{2}}{{N}_{e}}(\omega )$ is a spectrum of population fluctuations
\beq
	\left\langle \delta {{{\hat{N}}}_{e}}(\omega )\delta {{{\hat{N}}}_{e}}(\omega ') \right\rangle ={{\delta }^{2}}{{N}_{e}}(\omega )\delta (\omega +\omega '). 	\lb{sp_pop_fl}
\eeq
The field spectrum ${{n}_{1}}(\omega )$, $\left\langle \hat{a}_{1}^{+}(\omega ){{{\hat{a}}}_{1}}(\omega ') \right\rangle ={{n}_{1}}(\omega )\delta (\omega +\omega ')$, can be represented, with the help of  Eq.~\rf{a_1}, as
\beq
{{n}_{1}}(\omega )={{n}_{0}}(\omega )+n_{sp}(\omega)+n_{st}(\omega). \lb{n_1_sp}
\eeq
Here $n_0(\omega)$, given by Eq.~\rf{n0_sp}, is caused by the vacuum fluctuations of the lasing mode and the active medium polarization;  
\beq
        n_{sp}(\omega) = \frac{\kappa\gamma_{\perp}^2}{2N_{th}}\frac{N_1(\omega)}{S(\omega)} \lb{n_sp_0}    
\eeq
is due to the effect of the population fluctuations on spontaneous emission: we see that $n_{sp}(\omega)$ does not depend explicitly on the mean photon number;
\beq
        n_{st}(\omega) = \left(\frac{\kappa\gamma_{\perp}}{N_{th}}\right)^2\frac{\left( {{n}_{0}}*{{\delta }^{2}}{{N}_{e}} \right)_{\omega}}{S(\omega)} \lb{n_st_0}    
\eeq
is proportional to the mean photon number $n_0$, appeared in $\left( {{n}_{0}}*{{\delta }^{2}}{{N}_{e}} \right)_{\omega}$ and, therefore, it is due to the effect of the population fluctuations on the stimulated emission.

Replacing $\left( {{n}_{0}}*{{\delta }^{2}}{{N}_{e}} \right)_{\omega}$  by $n_0\delta^2N_e(\omega)$ in \rf{n_st_0} we come to the approach of \ct{Protsenko_2021}, which is good, if the field spectrum $n_0(\omega)$ is much narrower than the population fluctuation spectrum $\delta^2N_e(\omega)$. This is true for the high excitation, when the laser generate coherent radiation, so $n_0(\omega)\approx n_0\delta(\omega)$ where $\delta(\omega)$ is Dirac delta-function. The term $n_{sp}(\omega)$ does not appear in the approach of \ct{Protsenko_2021}, which does not take into account the influence of population fluctuations on the spontaneous emission into the lasing mode. 

With the derivation of Eqs.~\rf{n_1_sp} -- \rf{n_st_0} we suppose, that  ${{\hat{a}}_{0}}\delta {{\hat{N}}_{e}}$, in the first-order approximation, is  not correlated with ${{\hat{F}}_{a}}$ and $\hat{F}_{v}^{(1)}$.

We find ${{N}_{1}}(\omega )$ demanding  Bose commutation relations $\left\langle \left[ {{{\hat{a}}}_{1}},\hat{a}_{1}^{+} \right] \right\rangle =1$. From Eq.~\rf{a_1} we obtain
\beqr
	&{{\left[ {{{\hat{a}}}_{1}},\hat{a}_{1}^{+} \right]}_{\omega }}={{\left[ {{{\hat{a}}}_{0}},\hat{a}_{0}^{+} \right]}_{\omega }}+&\lb{com_a1}\\ &\displaystyle  \{{{(\kappa {{\gamma }_{\bot }}/{{N}_{th}})}^{2}}{{\left( \left[ {{{\hat{a}}}_{0}},\hat{a}_{0}^{+} \right]*{{\delta }^{2}}{{N}_{e}} \right)}_{\omega }}-\kappa \gamma _{\bot }^{2}{{N}_{1}}(\omega )/{{N}_{th}}\}/{S(\omega )},& \nonumber
\eeqr
with the spectrum $\left[ {{{\hat{a}}}_{0}},\hat{a}_{0}^{+} \right]_{\omega}$ given by Eq.~\rf{sp_com}. We know that ${{(2\pi )}^{-1}}\int\limits_{-\infty }^{\infty }{{{\left[ {{{\hat{a}}}_{0}},\hat{a}_{0}^{+} \right]}_{\omega }}}d\omega =1$. Therefore  ${{(2\pi )}^{-1}}\int\limits_{-\infty }^{\infty }{{{\left[ {{{\hat{a}}}_{1}},\hat{a}_{1}^{+} \right]}_{\omega }}}d\omega =1$, if the nominator in the second term on the right in Eq.\rf{com_a1} is zero, which is true when
\beq
	{{N}_{1}}(\omega )=(\kappa /{{N}_{th}}){{\left( \left[ {{{\hat{a}}}_{0}},\hat{a}_{0}^{+} \right]*{{\delta }^{2}}{{N}_{e}} \right)}_{\omega }}. \lb{N1_expr}
\eeq
Incerting ${{N}_{1}}(\omega )$ from Eq.~\rf{N1_expr} into Eq.~\rf{n_sp_0} we find
\beq
        n_{sp}(\omega) = \left(\frac{\kappa\gamma_{\perp}}{N_{th}}\right)^2\frac{{{\left( \left[ {{{\hat{a}}}_{0}},\hat{a}_{0}^{+} \right]/2 *{{\delta }^{2}}{{N}_{e}} \right)}_{\omega }}}{S(\omega )}. 	\lb{n1_sp}
\eeq
We see that  $n_{sp}(\omega)$  depends on the convolution of the population fluctuation spectrum $\delta^2N_e(\omega)$ with the spontaneous emission noise   spectrum. Indeed, the spectrum ${{\left[ {{{\hat{a}}}_{0}},\hat{a}_{0}^{+} \right]}_{\omega }}/2$, in the convolution in Eq.~\rf{n1_sp}, is a spectrum of vacuum field  fluctuations in the lasing mode, or a “spectrum of the half of a photon”: ${{(2\pi )}^{-1}}\int\limits_{-\infty }^{\infty }{\left( {{\left[ {{{\hat{a}}}_{0}},\hat{a}_{0}^{+} \right]}_{\omega }}/2 \right)}d\omega =1/2$. 

In order to find $n_{sp}(\omega)$ and  $n_{st}(\omega)$ we must know the spectrum of population fluctuations $\delta^2N_e(\omega)$.
From Eq.\rf{FC_3} we find $\delta\hat{N}_e(\omega)$ and the population fluctuation spectrum
\beq
\delta^2{N}_e(\omega) = \frac{\Omega^2\delta^2{\Sigma}(\omega)+2D_{N_eN_e}}{\omega^2+\gamma_P^2}, \lb{pop_fl_sp1a}
\eeq
where $\delta^2{\Sigma}(\omega)$ is the spectrum of $\delta\hat{\Sigma}(\omega)$. With calculations of $\delta^2{N}_e(\omega)$ we  use the same approximation as in \ct{Protsenko_2021} neglecting by correlations between polarization and population fluctuations, i.e. between $\hat{F}_v$ and $\hat{F}_{N_e}$, which is good approximation at a large number of emitters $N_0\gg 1$. Diffusion coefficient $2D_{N_eN_e} = \gamma_{\parallel}(PN_g+N_e)$ is the same as in the rate equation laser theory \ct{Coldren}.  

We find $\delta\hat{\Sigma}(\omega)$ from Eqs.~\rf{binary_eqs_fl} written Appendix~\ref{Sigma_eqs}  in the zero-order approximation on $\delta\hat{N}_e$. Then we find the spectrum $\delta^2{\Sigma}(\omega)$ from Eq.~\rf{Sigma_spect}. Explicit expression for $\delta^2{\Sigma}(\omega)$ is cumbersome, so we do not present it here.
With $\delta^2{\Sigma}(\omega)$ we integrate the spectrum \rf{pop_fl_sp1a} over frequencies and find the population fluctuation dispersion $\delta^2{N}_e$.

%
\begin{figure}[thb]\bc
\includegraphics[width=8cm]{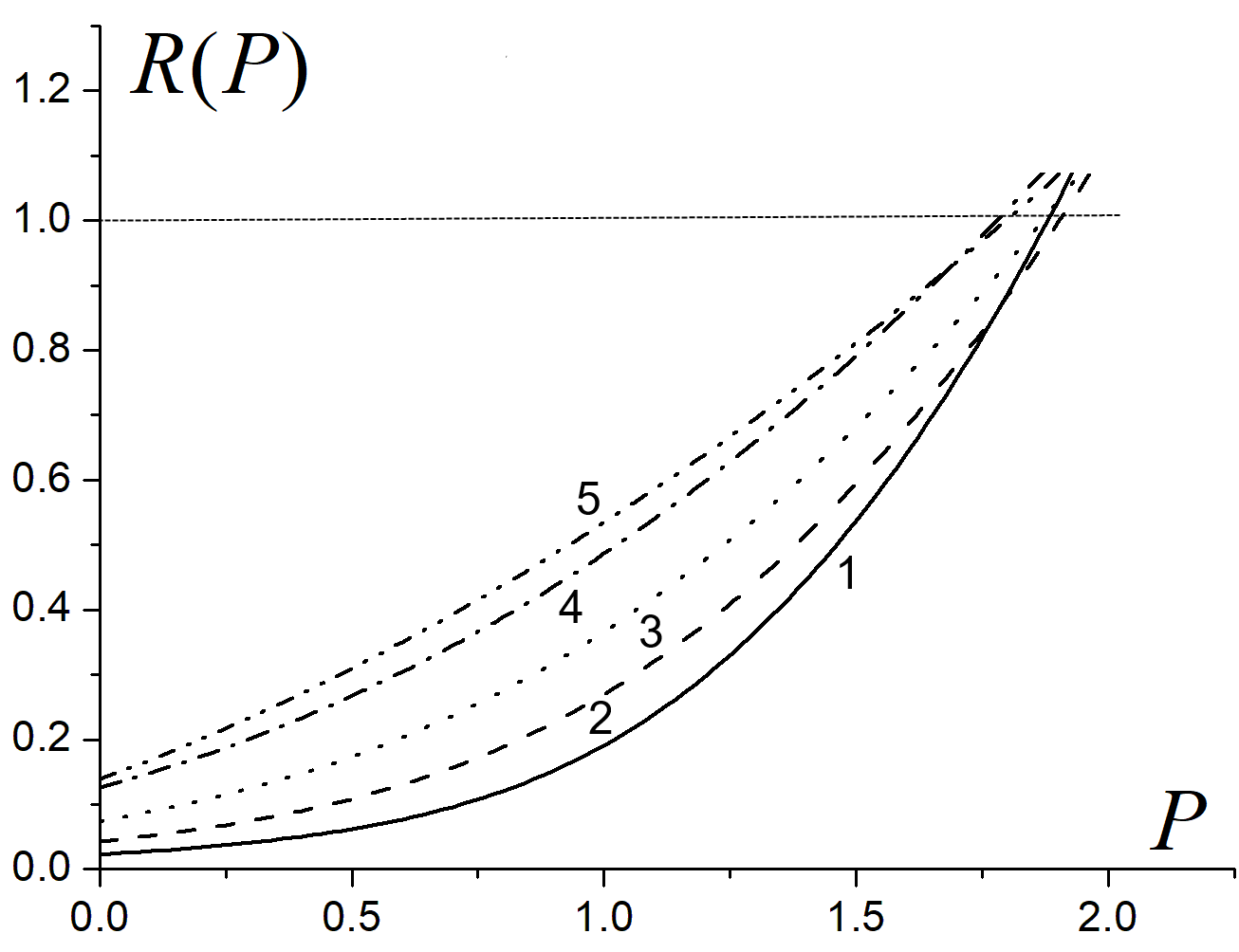}
\caption{The relative difference $R(P)$ of the population fluctuation dispersion found with and without $\delta\hat{\Sigma}$ for $\gamma_{\perp} = 5$ (curve 1), $10$ (2), $20$ (3), $50$ (4) and $500$ (5). $R(P)<1$, so  population fluctuations caused by $\delta\hat{\Sigma}$ (the first term in Eq.~\rf{pop_fl_sp1a}) is smaller than population fluctuations caused by the second term in Eq.~\rf{pop_fl_sp1a}  at the weak excitation, when the pump rate $P<2$. }
\label{Fig2x}\ec
\end{figure}
%
Fig.~\ref{Fig2x} shows the relative difference
\beq
R=\delta^2{N}_e/\delta^2{N}_e^{(0)}-1 \lb{rel_disp}
\eeq
of  $\delta^2{N}_e(P)$ found with the help of Eq.~\rf{pop_fl_sp1a}  and the population fluctuation dispersion $\delta^2{N}_e^{(0)}(P) = 2D_{N_eN_e}/2\gamma_P$ found by integrating Eq.~\rf{pop_fl_sp1a} without $\delta^2{\Sigma}(\omega)$.   
We see from Fig.~\ref{Fig2x} that $R<1$, which means that the contribution from 
$\delta\hat{\Sigma}$ to population    
is relatively small for $P<2$. So, for the sake of simplicity,  we drop the first term in 
 Eq.~\rf{FC_3} at the low excitation and approximate
\beq
        \delta\hat{N}_e(\omega) \approx \hat{F}_{N_e}(\omega)/(i\omega-\gamma_P). \lb{pop_fl_eq}
\eeq
Calculations based on the approximation \rf{pop_fl_eq} demonstrate  our method in a simplified setting,  however  approximation \rf{pop_fl_eq} is not a necessary part of the method.  
Approximation \rf{pop_fl_eq} considerably simplifies the calculation of convolutions in Eqs.~\rf{n_st_0} and \rf{n1_sp} and, in the meanwhile, shows, as we will see, the non-negligible influence of population fluctuations on the lasing at the low excitation. Straightforward but  cumbersome calculations of convolutions beyond the approximation \rf{pop_fl_eq} can be done with $\delta\hat{N}_e(\omega)$ satisfying Eq.~\rf{FC_3} and found from equations \rf{binary_eqs_fl} of  Appendix~\ref{Sigma_eqs}. We leave such calculations for the future.  

With the approximation \rf{pop_fl_eq} the spectrum of population fluctuations is
\beq
\delta^2N_e(\omega) = 2D_{N_eN_e}/(\omega^2+\gamma_P^2). \lb{pop_fl_sp}
\eeq
The mean photon number ${{n}_{1}}={{(2\pi )}^{-1}}\int\limits_{-\infty }^{\infty }{{{n}_{1}}(\omega )}d\omega$ depends on the mean population $N_e$ of the upper lasing states. $N_e$ can be found from the energy conservation law \rf{eql_1} with $n=n_1(N_e)$. 
\section{\label{Sec5}Results and discussion}
In examples we present results of calculations with parameters:  the  wavelength of the lasing transition $\lambda_0 = 1.55$~$\mu$m,  the background refractive index $n_r = 3.3$, the cavity mode volume $V_c = 10(\lambda_0/n_r)^3$ with $N_0 = 100$ emitters; a population relaxation rate $\gamma_{\parallel} = 10^9~s^{-1}$; the vacuum Rabi frequency $\Omega = (d/n_r)[\omega_0/(\varepsilon_0\hbar V_c)]^{1/2}$ with a dipole moment of the lasing transition $d = 10^{-28}$~Cm so that $\Omega = 34\gamma_{\parallel}$; the average atom-lasing mode-coupling factor $f=1/2$ and the cavity quality factor $Q = 1.2\cdot 10^4$ so $2\kappa = 100\gamma_{\parallel}$. 

We vary the dephasing rate $\gamma_{\perp}$ and the pump $P$  keeping all other parameters fixed.  $\gamma_{\perp}$ is varied between $50$~GHz ( $2\kappa/\gamma_{\perp} = 2$) to $1.5$~THz (with $2\kappa/\gamma_{\perp} = 0.07$). This is a realistic  region of  $\gamma_{\perp}$ for quantum dots~\ct{PhysRevB.46.15574}. We calculate the non-normalized $\beta$-factor  $\tilde{\beta}=g/\gamma_{\parallel}$ \ct{Protsenko_2021}, where  $g=4\Omega^2f/[\gamma_{\perp}(1+2\kappa/\gamma_{\perp})]$ is the spontaneous emission rate into the lasing mode and the rate $\gamma_{\parallel}$ includes all population losses in the lasing medium.  Within the chosen range for $\gamma_{\perp}$, $\tilde{\beta}$ varies from $15$  to $1.4$, so lasers with the chosen  parameters have significant amounts of spontaneous emission into the lasing mode.

Similar parameters can be found in photonic crystal nanolasers with  quantum-dot active media~\ct{doi:10.1063/1.5022958}; superradiant lasers with cold alkaline earth atoms~\ct{PhysRevX.6.011025, PhysRevA.96.013847,PhysRevA.81.033847,PhysRevA.98.063837},  rubidium atoms~\ct{Bohnet} and quantum dots~\ct{Jahnke}. 
These lasers are thresholdless, with a large non-normalized beta-factor and with  significant influence of collective effects (the superradiance)  \ct{Khanin,Belyanin_1998,Koch_2017,Andre:19,Protsenko_2021}. Population fluctuations in superradiant lasers are large \ct{Andre:19,Protsenko_2021}. We consider LED regime with relatively small dimensionless pump rate $P<2$, when the mean number of the cavity photons is of the order of one or less, and when the linewidth $\gamma_{las}$ of the lasing field is large $\gamma_{las} > \gamma_{\parallel}$.

\begin{figure}[thb]\bc
\includegraphics[width=8cm]{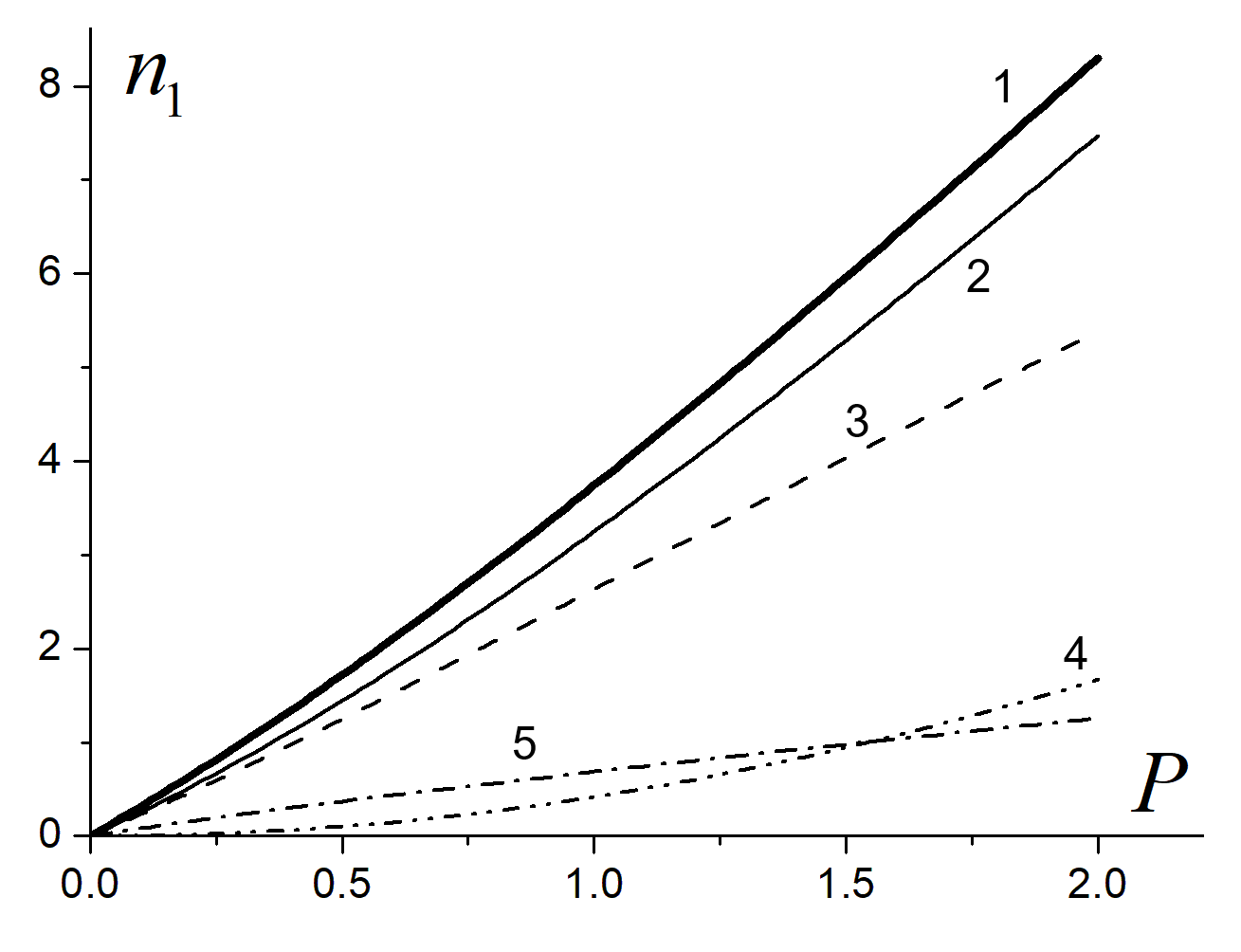}
\caption{The mean photon number $n_1$ versus the normalized pump rate $P$ for thresholdless superradiant laser with $2\kappa /{{\gamma }_{\bot }}=2$, ${{N}_{0}}=100$ resonant emitters, and non-normalised beta-factor \ct{Protsenko_2021} $\tilde{\beta} =15.4\gg 1$.  Curves 1 and 2 are found  with and without population fluctuations, respectively. $n_1$ in the curve 1 is the sum of values in curves 3, 4 and 5 taken with the same $P$ and population inversion $N$. The curve 3 is due to vacuum fluctuations in the lasing mode; curves 4 and 5 are contributions  of the effect of population fluctuations on spontaneous and on stimulated emission, correspondingly. The mean population inversion for curves 1, 3, 4 and 5 is smaller than for the curve 2 because population fluctuations accelerate the radiation and reduce the population inversion.}
\label{Fig3x}\ec
\end{figure}

The mean photon number $n_1(P)$ for $\gamma_{\perp} = 50\gamma_{\parallel}$ is shown in Fig.~\ref{Fig3x}, where we note the influence of population fluctuations on the lasing field. 
%
%
%
In Fig.~\ref{Fig3x} the bold solid curve 1 is $n_1(P)$ found in the first-order approximation  with population fluctuations. The thin solid curve 2 is $n_0$ found  without population fluctuations. The other curves are parts of $n_1$: the  curve 3 is due to fluctuations of polarization  with the spectrum $n_0(\omega)$ in Eq.~\rf{n_1_sp}; the curve 4 and the curve 5 are due to the effect of population fluctuations on spontaneous and on stimulated emission respectively, they are the integrals of spectra $n_{sp}(\omega)$ and $n_{st}(\omega)$ in Eq.~\rf{n_1_sp} correspondingly. The curve 1 is the sum of curves 3, 4 and 5, they depend on the same mean population inversion $N$ found from the energy conservation law \rf{eql_1}.   

We see in Fig.~\ref{Fig3x} that population fluctuations (curves 4 and 5) give a noticeable contribution into the mean cavity photon number (the curve 1). Comparing curves 1 and 2 in Fig.~\ref{Fig3x} we see that population fluctuations at the low excitation make a larger influence on the mean photon number than it was predicted with the standard perturbation approach used in \ct{Protsenko_2021}. In Fig.5 of \ct{Protsenko_2021} we see that $n$ found with and without population fluctuations almost coincide. This is because of the standard perturbation approach does not consider the influence of population fluctuations on spontaneous emission.

\begin{figure}[thb]\bc
\includegraphics[width=8cm]{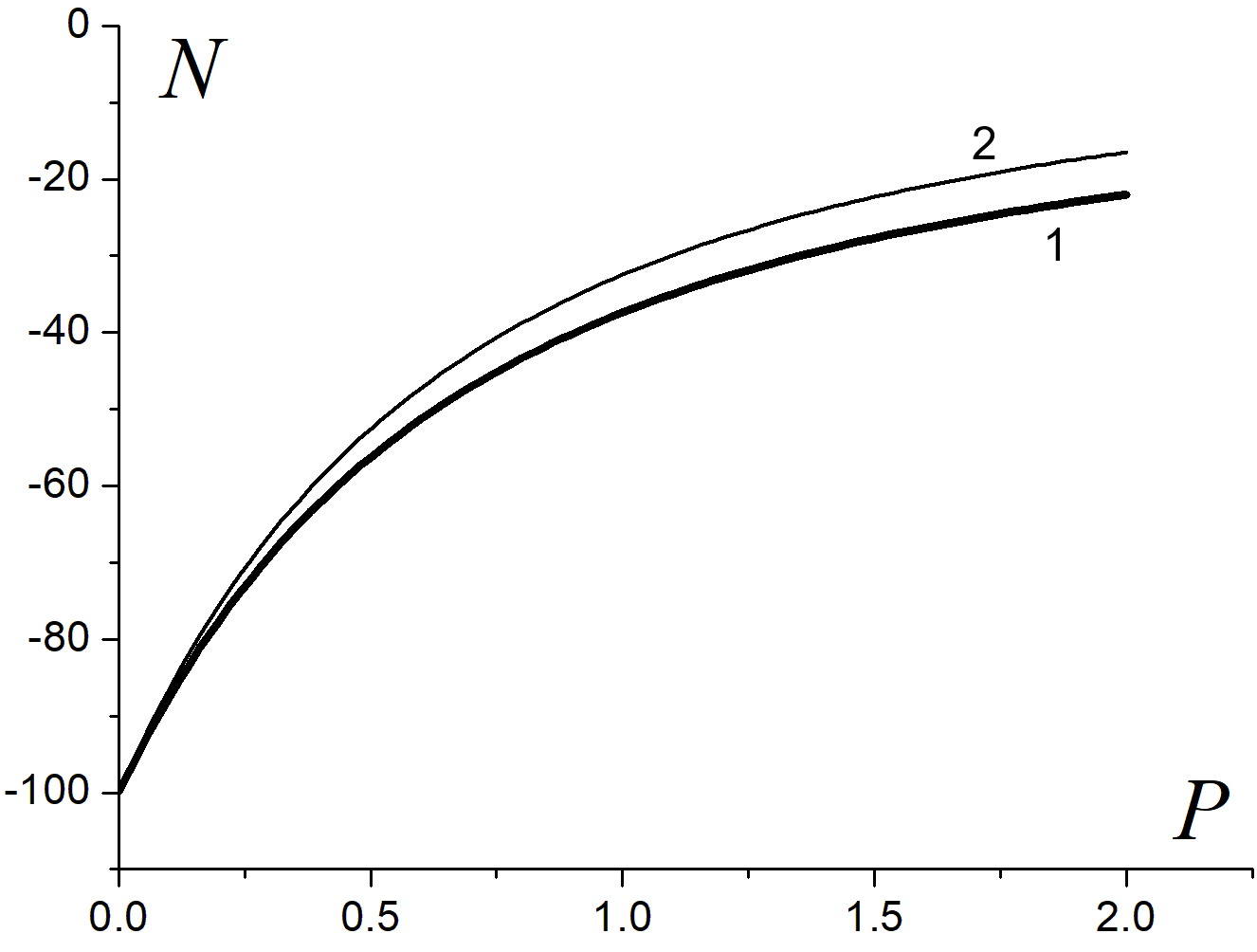}
\caption{The mean population inversion calculated with (curve 1) and without (curve 2) population fluctuations. Population fluctuations increase the radiation rate and deplete the population inversion. This is why the curve 1 goes below the curve 2.}
\label{Fig4x}\ec
\end{figure}

One can find that the population inversion $N$ for the curve 2 is {\em larger} than for curves 1, 3, 4 and 5, since $N$ is depleted, because of population fluctuations increase the radiation rate, see population inversions for curves 1 (with population fluctuations) and 2 (without population fluctuations) in Fig.~\ref{Fig4x}. This is why the curve 3 goes below the curve 2 in Fig.~\ref{Fig3x}. 

It is well-known that the spontaneous emission is stimulated by the vacuum fluctuations of the electromagnetic field \ct{Scully} and that a high  density of states of the field increases the spontaneous emission rate in the cavity (Pursell effect) \ct{PhysRev.69.674}. As an important finding we see that the population fluctuations increase the spontaneous (and the stimulated) emission rates into the lasing mode. Such emission rate increase may be important for highly efficient LEDs. We will estimate how large such increase can be. 

\begin{figure}[thb]\bc
\includegraphics[width=8cm]{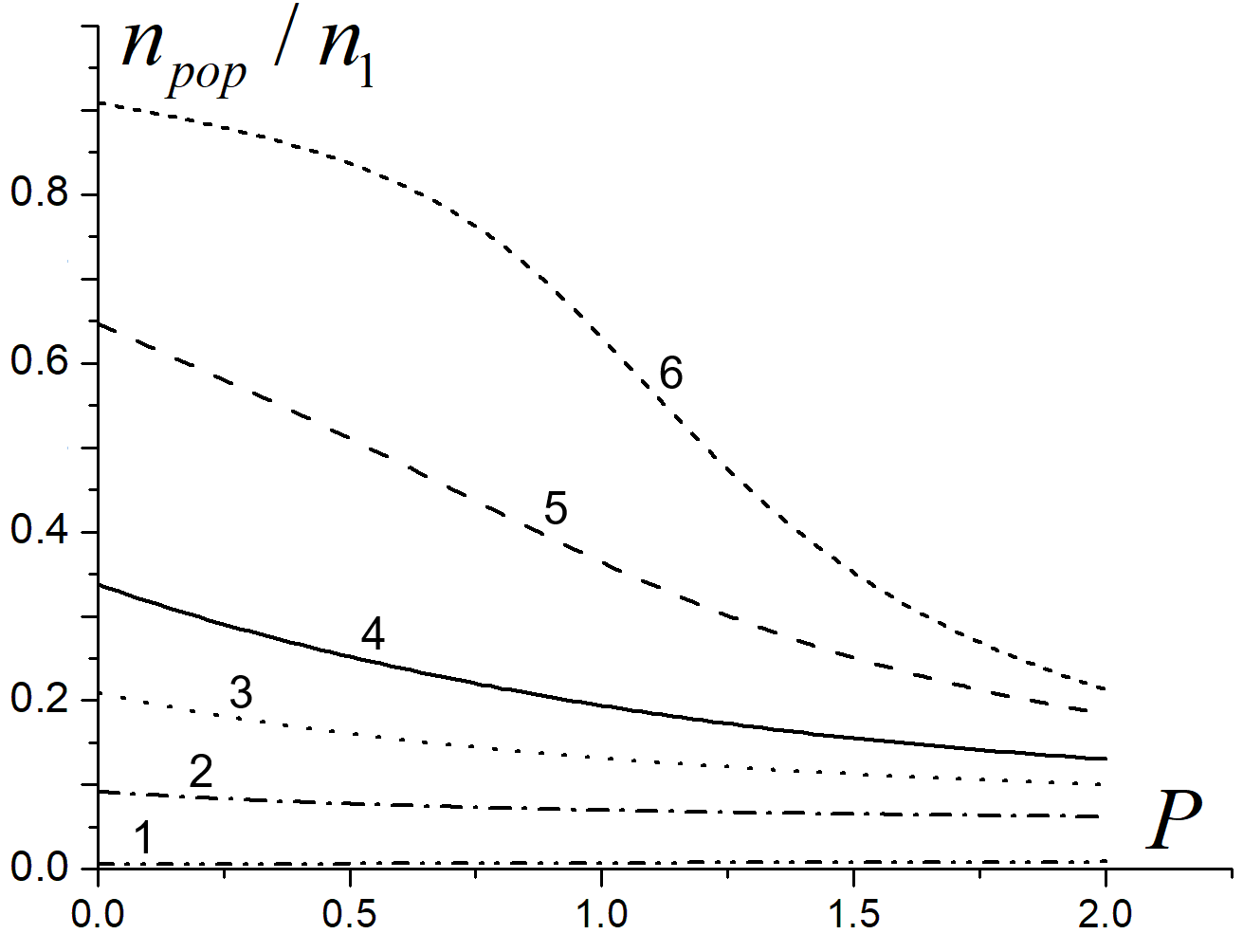}
\caption{The relative contribution of population fluctuations to the mean photon number for $\gamma_{\perp}/\gamma_{\parallel} = 1500$ (curve 1), $100$ (2), $50$ (3), $30$ (4), $10$ (5) and $2$ (6) and other parameters the same as for Fig.~\ref{Fig6x}. We see that for small pump almost all photons in the lasing mode are related with population fluctuations at small $\gamma_{\perp}$  approaching $\gamma_{\parallel}$  as for curves 5 and 6.}
\label{Fig5x}\ec
\end{figure}

We note in Fig~\ref{Fig3x}, that the contribution of population fluctuations into  spontaneous emission (the curve 4) dominates the contribution into stimulated emission (the curve 5) at weak pump $P<1.5$, when the cavity photon number is small. 
%
%
%
We introduce the  characteristic of the influence of the population fluctuations on the emission rate. For that we calculate the part $n_{pop}$ of the mean number of photons
\beq
n_{pop} = \frac{1}{2\pi}\int_{-\infty}^{-\infty}[n_{sp}(\omega)+n_{st}(\omega)]d\omega, \lb{Delta_n}
\eeq
caused by population fluctuations. Eq.~\rf{Delta_n} it is the sum of curves 4 and 5 in Fig~\ref{Fig3x}. The ratio $n_{pop}/n_1$  characterises the contribution of population fluctuations into the emission rates. Smaller $n_{pop}/n_1$ corresponds to a smaller influence of the population fluctuations. 
$n_{pop}/n_1$ is shown in Fig~\ref{Fig5x}
as a function of the pump $P$ for different $\gamma_{\perp}$. We see that $n_{pop}/n_1$ is reduced with $P$ and grows for smaller $\gamma_{\perp}$. 
For curves 5 and 6 $n_{pop}/n_1$ is close to 1, which means that almost all photons in the lasing mode are related with population fluctuations, when $P\rightarrow 0$ and for small $\gamma_{\perp}\rightarrow \gamma_{\parallel}\ll 2\kappa$. Thus we  conclude that population fluctuations may considerably increase the emission rate at a weak pump in lasers with a narrow lasing transitions such that $\gamma_{\perp}\ll 2\kappa$.  In such lasers population fluctuations are high and collective effects are significant \ct{Protsenko_2021}.

%
%
The limit of $n_{pop}/n_1$ close to 1, however, does not correspond to the perturbation approach on population fluctuations, 
so curves 5 and 6 in Fig.~\ref{Fig5x} must be re-considered
in higher orders of the approximation. We show curves 5 and 6 in  Fig.~\ref{Fig5x} since they display a trend of the increase of the emission rate by population fluctuations, when (a) the pump $P$ became smaller, and  (b) for bad-cavity lasers, where the cavity dumping rate $2\kappa$ is relatively large $2\kappa >\gamma_{\perp}$. Fig.~\ref{Fig5x} indicates a possibly of a high acceleration of the radiation from LEDs at a weak pump and on corresponding increase of the LED efficiency by population fluctuations. Determining the maximum radiation rate increase at the weak pump, is an interesting topic important for applications, but it is beyond the first-order perturbative scheme. We leave this topic for the future.  From Fig.~\ref{Fig5x} we learn, that the expected increase of the radiation rate by population fluctuations may be of the order, or even larger, that the radiation rate taken without population fluctuations.      
 
%
\begin{figure}[thb]\bc
\includegraphics[width=8cm]{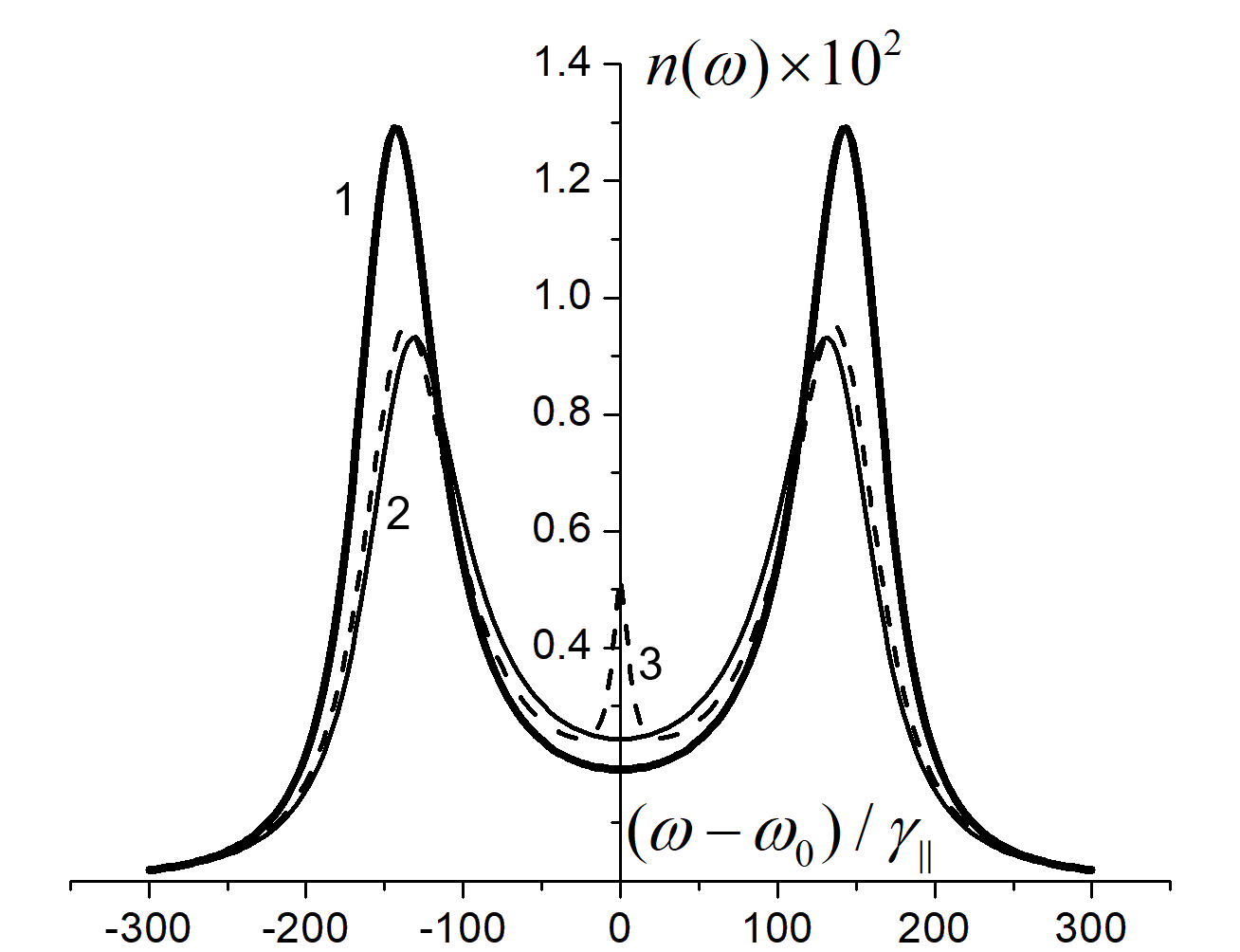}
\caption{Photon number spectra found with (the solid curve 1) and without (the thin curve 2) population fluctuations. $P=1$, other parameters are the same as for Fig.~\ref{Fig3x}. The  dashed curve 3 is a result of \ct{Protsenko_2021} found with population fluctuations.   The narrow peak in the center of the curve 3 disappears in present approach, while the mean photon number (the height of the spectrum) increases -- compare curves 1 and 3.  }
\label{Fig6x}\ec
\end{figure}
%

Fig.~\ref{Fig6x} shows  spectra of the lasing field calculated with (the solid curve 1) and without (the thin curve 2) population fluctuations for $\gamma_{\perp} = 50\gamma_{\parallel}$ (the same as for Fig.~\ref{Fig3x}) and for  $P=1$. Two peaks in spectra in Fig.~\ref{Fig6x} are because of the collective Rabi splitting \ct{Andre:19}. 

According with Fig.~\ref{Fig6x}, present approach  does not predict a narrow peak in the center of spectra found in \ct{Protsenko_2021}. Instead we see the increase of sideband peaks due to population fluctuations. This is because of the approximation $(\hat{a}\delta\hat{N}_e)_{\omega}\approx\sqrt{n}\delta\hat{N}_e(\omega) $ used in \ct{Protsenko_2021} ignores the finite width of the field spectrum and the effect of population fluctuations on the spontaneous emission into the lasing mode. It is not appropriate at the low excitation in the bad cavity lasers, where  population and the field fluctuations are large. 

Thus we correct results of \ct{Protsenko_2021} for the LED regime by making more accurate description of population fluctuations. Here  we use a convolution of spectra for calculating nonlinear terms in laser HLE  and corrected diffusion coefficients, while in  \ct{Protsenko_2021}  the approach for a high-excitation regime was directly extended to the low-excitation LED regime.     
%
%
\begin{figure}[thb]\bc
\includegraphics[width=8cm]{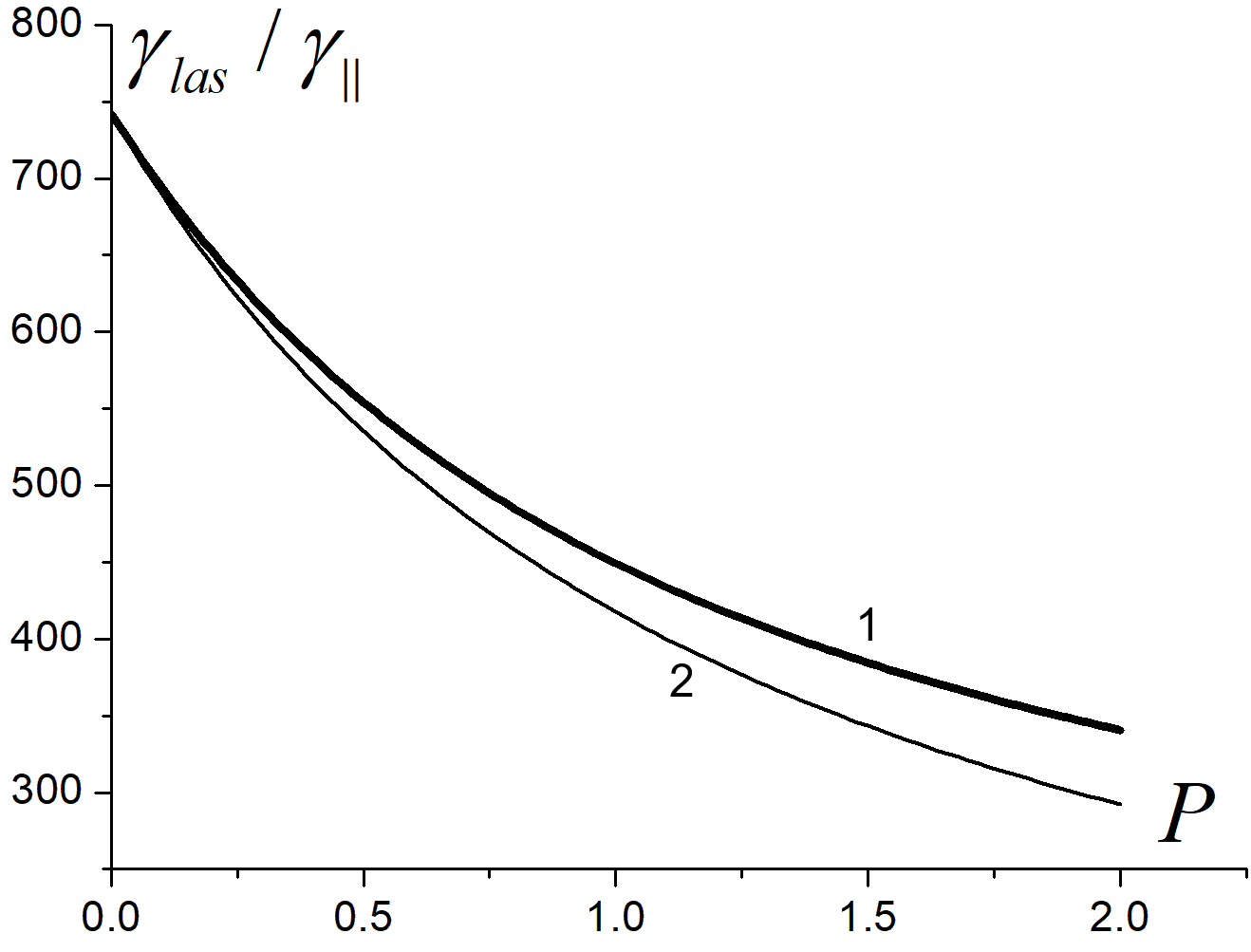}
\caption{Laser linewidth with (the solid curve 1) and without  (the thin curve 2) population fluctuations for the same parameters as for Fig.~\ref{Fig3x}.}
\label{Fig7x}\ec
\end{figure}
%

Fig.~\ref{Fig7x} shows the laser linewidth \ct{Protsenko_2021}
\beq
\gamma_{\rm las} = \frac{2\kappa+\gamma_{\perp}}{{\sqrt{2}}}\left\{r-1+\sqrt{(r-1)^2+r^2}\right\}^{1/2},
\eeq\[r=\frac{4\kappa\gamma_{\perp}}{(2\kappa+\gamma_{\perp})^2}(1-N/N_{\rm th}),\]
found with (the curve 1) and without (the curve 2) population fluctuations. The linewidth of the laser, with population fluctuations taken into account, is larger than the linewidth of the laser where population fluctuations are neglected, so population fluctuations broad the lasing spectrum.

\section{Conclusion}

We consider population fluctuations as a perturbation in quantum nonlinear stochastic equations for the laser  and present an approximate approach for solving such equations analytically in various orders on perturbations. As an example, we consider Maxwell-Bloch  equations for the laser in the low-excitation (or LED) regime. Spectra of nonlinear terms  are found as convolutions of spectra calculated in the zero-order approximation, when population fluctuations are neglected. This approach improve the method of \ct{Protsenko_2021}, where nonlinear terms have been linearized around mean values, which is not an accurate approximation at the low excitation. Diffusion coefficients for Langevin forces  are found from the requirement, that Bose commutation relations for operators of the lasing field are preserved. 

We found that population fluctuations accelerate spontaneous and stimulated emissions, increase the radiation rate and, as a consequence, the mean number of lasing photons. Population fluctuations broad the lasing spectrum. We found larger mean photon number at the low excitation and the absence of small peaks in the center of the field spectrum shown in \ct{Protsenko_2021} and correct results of \ct{Protsenko_2021}.

Population fluctuations are high in bad cavity lasers with large gain and relatively narrow lasing transitions, such as superradiant lasers, where collective effects are significant. A large part of the radiation in LED regime in such lasers  may be related with the population fluctuations. 

 Lasers or LEDs with the radiation rate, increased by population fluctuations, may find applications as miniature and efficient broadband light sources. 

Our approach may be applied for theoretical analysis of various resonant systems in nonlinear and quantum optics as, for example, optical parametric oscillator in the cavity \ct{Wang2021}.

\begin{acknowledgments}
We wish to acknowledge the stimulated discussions  the notes and advises from  porfessor Jesper M$\ddot{\text o}$rk and professor Martijn Wubs from Photonic department of the Danish Technical university.
\end{acknowledgments}

\appendix

\section{\label{App_op}Fourier-expansion for operators}

We consider  Fourier-expansion  of  Bose-operator $\hat{a}(t){{e}^{-i{{\omega }_{0}}t}}$ of the lasing mode, where $\hat{a}(t)$  is changed much slowly than ${{e}^{-i{{\omega }_{0}}t}}$. 

In the case of classical field complex amplitude  $a(t)$  can be represented as Fourier-integral
\beq
a(t)=\frac{1}{\sqrt{2\pi }}\int_{-\infty }^{\infty }{a(\omega )}{{e}^{-i\omega t}}d\omega, \lb{1}
\eeq
where $a(\omega )$ is Fourier-component of $a(t)$. Expression \rf{1} describes the  physical fact, that the electromagnetic field is a superposition of monochromatic components of different frequencies \ct{Jackson:100964}. According with Heisenberg correspondence principle \ct{HCP}  Fourier-expansion~\rf{1} remains true  for quantum electromagnetic field, so  classical variables in Eq.~\rf{1} can be replaced by operators  
\beq
\hat{a}(t)=\frac{1}{\sqrt{2\pi }}\int_{-\infty }^{\infty }{\hat{a}(\omega )}{{e}^{-i\omega t}}d\omega. \lb{2}
\eeq
We will come to Eq.~\rf{2}  another way, by a transition from Schrodinger to Heisenberg operators with the help of the evolution operator \ct{doi:10.1142/S0129055X07003206}. 

Suppose $\left| \Psi  \right\rangle $ is a wave function of the system (of the laser in our case) and of baths interacting with the system. $\left| \Psi  \right\rangle $ is, therefore, the eigenfunction of Hamiltonian $H$  of the system and baths. In Heisenberg representation $\left| \Psi  \right\rangle $ does not depend on time. We average the operator   $\hat{a}$ over $\left| \Psi  \right\rangle $ 
\beq
    \left< \Psi  \right|\hat{a}(t) \left| \Psi  \right\rangle = a(t). \lb{3}
\eeq
$a(t)$ is a random function of time, because of quantum fluctuations of the lasing mode and fluctuations due to the interaction of the mode with baths.  In the stationary case $a(t)$ corresponds to the stationary random process. 

Operator $\hat{a}(t)$ is related with the time-independent Schrodinger operator $\hat{a}_{sh}$ by the transformation
\beq
        \hat{a}(t)=\exp \left( iHt/\hbar  \right){{\hat{a}}_{Sh}}\exp \left( -iHt/\hbar  \right), \lb{4}
\eeq
where $\exp \left( -iHt/\hbar  \right)$ is the evolution operator \ct{Landau_Quant}. 

Suppose, for simplicity, that $\left| \Psi  \right\rangle $ can be expanded over states with discreet spectrum, %
\beq
\left| \Psi  \right\rangle =\sum\limits_{i=1}^{\infty }{\left| {{\Psi }_{i}} \right\rangle }, \lb{expand0}
\eeq
where $\left\{ \left| {{\Psi }_{i}} \right\rangle  \right\}$ 
is a complete set of mutually orthogonal eigenstates of Hamiltonian $H$. 

We  take a unity operator $\hat{1}$   \ct{Andrews:20,Cohen-Tannoudji:101367}   
\beq
\hat{1}=\sum\limits_{i=1}^{\infty }{\left| {{\Psi }_{i}} \right\rangle \left\langle  {{\Psi }_{i}} \right|}, \lb{5}
\eeq
and insert $\hat{1}$ into Eq.~\rf{4} on the right and on the left side to the operator ${\hat{a}}_{Sh}$. After this we average Eq.~\rf{4} over the state $\left| \Psi  \right\rangle $ and come to
\beq
  a(t)=\sum\limits_{i,j=1}^{\infty }{\left\langle  \Psi  \right|e^{ iHt/\hbar}\left| {{\Psi }_{i}} \right\rangle }{{a}_{ij}}\left\langle  {{\Psi }_{j}} \right|e^{-iHt/\hbar}\left| \Psi  \right\rangle, \lb{6}  
\eeq
where ${{a}_{ij}}=\left\langle  {{\Psi }_{i}} \right|{{\hat{a}}_{Sh}}\left| {{\Psi }_{j}} \right\rangle$ is a matrix element of the operator ${\hat{a}}_{Sh}$. $\left| {{\Psi }_{i}} \right\rangle$ are eigenfunctions of Hamiltonian $H$, $\left| {{\Psi }} \right\rangle$ is a superposition of states $\left| {{\Psi }_{i}} \right\rangle$, therefore 
\beq
\left\langle  \Psi  \right|e^{ iHt/\hbar }\left| {{\Psi }_{i}} \right\rangle =e^{i{{E}_{i}}t/\hbar}, \hspace{0.25cm} \left\langle  {{\Psi }_{j}} \right|e^{-iHt/\hbar}\left| \Psi  \right\rangle =e^{ -i{{E}_{j}}t/\hbar}, \lb{7}
\eeq
where ${E}_{i}$ is the energy of the state $\left| {{\Psi }_{i}} \right\rangle$. We insert Eqs.~\rf{7} into Eq.~\rf{6} and come to
\beq
a(t)=\sum\limits_{i,j=0}^{\infty }{{{a}_{ij}}{{e}^{-i{{\omega }_{ij}}t}}}, \lb{8}
\eeq
where ${{\omega }_{ij}}=\left( {{E}_{i}}-{{E}_{j}} \right)/\hbar $.

We consider resonant systems, where the most populated states have the energy close to $\hbar\omega_0$, so ${\omega }_{ij}\ll\omega_0$. Then we assume  that matrix elements $a_{ij}$ depend only on ${{E}_{i}}-{{E}_{j}}$, but not on ${{E}_{i}}$ or ${{E}_{j}}$ separately. Precisely, the dependence on ${{E}_{i}}\approx{{E}_{j}}$ is the same for relevant matrix elements  taken into account. Therefore $a_{ij} = a(\omega_{ij})$. We re-arrange terms $a_{ij}e^{-i\omega_{ij}t}$ in the sum \rf{8} in the ascending order on $\omega_{ij}$, use the  index $k$ instead of two indexes $i$ and $j$ and re-write Eq.~\rf{8} as the sum over $k$ 
\beq
a(t)=\sum_{k=0}^{\infty }a(\omega_k)e^{-i\omega_kt}. \lb{9}
\eeq
Eq.~\rf{9} relates the mean $a(t)$ and matrix elements $a(\omega_k)$ of  Schredinger operator $\hat{a}_{Sh}$. Matrix elements $a(\omega_k)$  define the operator $\hat{a}(\omega_k)$, so we can rewrite the relation \rf{9} in terms of operators
\beq
\hat{a}(t)=\sum_{k=0}^{\infty }\hat{a}(\omega_k)e^{-i\omega_kt}. \lb{10}
\eeq
Taking in Eq.~\rf{10} the limit of continues spectrum we come to Foruer-integral \rf{2} for the operator $\hat{a}(t)$.

From Eq.~\rf{4} we write
\beq
        {{\hat{a}}_{Sh}}=\exp \left(-iHt/\hbar  \right)\hat{a}(t)\exp \left(iHt/\hbar  \right). \lb{4_inv}
\eeq
Starting with Eq.~\rf{4_inv} we come to the reverse Fourier-transform
\beq
\hat{a}(\omega)=\frac{1}{\sqrt{2\pi }}\int_{-\infty }^{\infty }{\hat{a}(t )}{{e}^{i\omega t}}dt. \lb{2_inv}
\eeq
similar way as we come from Eq.~\rf{4} to Eq.~\rf{2}.

We  prefer to work with Foruer-expansions \rf{10} or \rf{2} for operators instead of the  mean values as Eq.~\rf{9}. Working with operators we can preserve commutation relations. The expansion \rf{9} for means neglects commutation relations. Obviously, that $a^*(t)a(t)=a(t)a^*(t)$ while ${{\hat{a}}^{+}}(t)\hat{a}(t)\ne \hat{a}(t){{\hat{a}}^{+}}(t)$. Preserving commutation relations for the field operators is important for correct description of fluctuations at small number of photons. 

We note, that there are a random function of time $a(t)$ on the left in Eq.~\rf{9} and a random function of frequency $a_{ij}(\omega)$  on the right in Eq.~\rf{9}.  A random set of frequencies $\omega_k$ corresponds to every realisation of the random process, described by $a(t)$. This way the correspondence between random processes in the time and in the frequency domains are established, for example, in numerical methods of generation of a random signal \ct{Kuzmenko}. Practically, at numerical calculations, $\omega_k$ may be chosen  homogeneously distributed over some interval $\left[ -{{\omega }_{\max }},{{\omega }_{\max }} \right]$, where ${\omega }_{\max }$ is something larger than the expected half of the maximum linewidth of spectra of the system \ct{Kuzmenko}.   

So each set of random frequencies corresponds to particular realization of the random process. Such a realization may be an analog of the path integral  \ct{PhysRevA.26.451,doi:10.1063/5.0055815}. Mean values of operators are the result of the averaging over many realizations.   

Mean values of Fourier-component operators, for example, $\left\langle \hat{a}(\omega )\delta {{{\hat{N}}}_{e}}(\omega ) \right\rangle $, are averaged over many realizations of the random processes with Fourier-expansion as Eq.~\rf{9}, where a  random set of frequencies is chosen for each realization.

\section{\label{AppB}Spectrum of the operator product}

It is sufficient to know power spectra in order to describe the system in the stationary state. Here we   calculate spectra of operator products approximately in the perturbation approach.

We carry out Fourier-expansion of the operator $\hat{a}^+$
\beq
\hat{a}^+(t)=\frac{1}{\sqrt{2\pi }}\int_{-\infty }^{\infty }{\hat{a}^+(-\omega )}{{e}^{-i\omega t}}d\omega\lb{11}
\eeq
and take the mean $\left\langle {{{\hat{a}}}^{+}}(t)\hat{a}(t+\tau ) \right\rangle$. In the stationary case  $\left\langle {{{\hat{a}}}^{+}}(t)\hat{a}(t+\tau ) \right\rangle$ does not depend on $t$. Therefore, if we write $\left\langle {{{\hat{a}}}^{+}}(t)\hat{a}(t+\tau ) \right\rangle$ with Fourier-expansions \rf{2} and \rf{11} 
\beq
\frac{1}{2\pi }\int\limits_{-\infty }^{\infty }{\left\langle {{{\hat{a}}}^{+}}(-\omega )\hat{a}(\omega ') \right\rangle }{{e}^{-i\left( \omega +\omega ' \right)t-i\omega '\tau }}d\omega d\omega ', \lb{12}
\eeq
it must be that 
\beq
    \left\langle \hat{a}^+(-\omega )\hat{a}(\omega ') \right\rangle =n(\omega)\delta(\omega+\omega'). \lb{13}
\eeq
Physical meaning of Eq.~\rf{13} is that there is no transitions from states of photons with  different energies and $\omega\neq\omega'$ in the stationary state: the probability of such transitions, proportional to $\left<\hat{a}(\omega)\hat{a}(\omega')\right>$, is zero. So the matrix of the operator $\hat{a}^+(\omega)\hat{a}(\omega')$ is diagonal in the stationary state, as well as matrices of  binary products of other Fourier-component operators. This fact  simplifies calculations.  

The mean number $n$ of photons in the lasing mode is 
\beq
    n = \left\langle {{{\hat{a}}}^{+}}(t)\hat{a}(t) \right\rangle = \frac{1}{2\pi}\int_{-\infty}^{\infty}n(\omega)d\omega, \lb{14}
\eeq
so $n(\omega)$ is a power spectrum  of the lasing field. 

We have seen, that $n(\omega)$ is a diagonal matrix element of the operator $\hat{a}^+(\omega)\hat{a}(\omega')$ in the basis $\left\{ \left| {{\Psi }_{i}} \right\rangle  \right\}$  of states of the laser and baths. Therefore
\beq
dp_n(\omega)=n(\omega)d\omega/(2\pi n) \lb{prob1}
\eeq
is a probability that the lasing field is in  states with energies in the interval from $\hbar(\omega_0+\omega)$ to $\hbar(\omega_0+\omega+d\omega)$. $n(\omega)/(2\pi n)$ is, therefore, a probability density. 

The binary product of Fourier-component operators $\delta \hat{N}_e(\omega)$ of population fluctuations is 
\beq
\left\langle \delta {{{\hat{N}}}_{e}}(\omega )\delta {{{\hat{N}}}_{e}}(\omega ') \right\rangle =\delta^2N_e(\omega )\delta (\omega +\omega '). \lb{15}
\eeq
Here we write ${{{\hat{N}}}_{e}}(\omega )$, not $\hat{N}_e^+(-\omega )$ (compare with Eq.~\rf{13}), because of population fluctuations are real quantities and $\delta \hat{N}_e^+(-\omega )=\delta \hat{N}_e(\omega )$.

We consider binary products $\hat{a}(t)\delta {{\hat{N}}_{e}}(t)$ and $\hat{a}^+(t)\delta {{\hat{N}}_{e}}(t)$ with zero mean $\left\langle \hat{a}\delta {{{\hat{N}}}_{e}} \right\rangle =0$. The fact, that such mean is zero follows from Eqs.~\rf{MBE_St}, when $\left<\hat{a}\right> = 0$ and $\left<\hat{v}\right> = 0$.  

Suppose, ${{S}_{a{{N}_{e}}}}(\omega )$ is the spectrum of the binary products of operators $\hat{a}\delta\hat{N}_e$ 
We  write, the same way as in Eq.~\rf{14},
\beq
\left\langle {{{\hat{a}}}^{+}}(t)\delta {{{\hat{N}}}_{e}}(t)\hat{a}(t)\delta {{{\hat{N}}}_{e}}(t) \right\rangle =\frac{1}{2\pi }\int\limits_{-\infty }^{\infty }{{{S}_{a{{N}_{e}}}}(\omega )d\omega }. \lb{16}
\eeq
We will show how $S_{aN_e}(\omega )$ is expressed through the lasing field spectrum $n(\omega)$ and the spectrum $\delta^2N_e(\omega )$ of the population fluctuations 
\beq
        \left\langle \delta \hat{N}_{e}^{2}(t) \right\rangle =\frac{1}{2\pi }\int\limits_{-\infty }^{\infty }\delta^2N_e(\omega )d\omega . \lb{17}
\eeq
In follows from the analysis in Appendix~\ref{App_op} that  Fourier-component operator is expressed through the time-dependent operator by the  Fourier-transform 
\beq
{{\left( \hat{a}\delta {{{\hat{N}}}_{e}} \right)}_{\omega }}=\frac{1}{2\pi }\int\limits_{-\infty }^{\infty }{\hat{a}(t)\delta {{{\hat{N}}}_{e}}(t){{e}^{i\omega t}}dt}.  \lb{18}
\eeq
Here $\left( \hat{a}\delta\hat{N}_e\right)_{\omega }$ is  Fourier-component of $\hat{a}(t)\delta \hat{N}_e(t)$. We insert  Fourier-expansions of $\hat{a}(t)$ and $\delta \hat{N}_e(t)$ into Eq.~\rf{18} and obtain 
\beq
 {{\left( \hat{a}\delta {{{\hat{N}}}_{e}} \right)}_{\omega }}=\int\limits_{-\infty }^{\infty }{\hat{a}({{\omega }_{1}})\delta {{{\hat{N}}}_{e}}}({{\omega }_{2}}){{e}^{-i\left( {{\omega }_{1}}+{{\omega }_{2}}-\omega  \right)t}}\frac{d\omega _1d\omega_2dt}{\left( 2\pi  \right)^{3/2}}.       \lb{19}
\eeq
We take the integral over the time in Eq.~\rf{19} using that
\beq
\frac{1}{2\pi }\int\limits_{-\infty }^{\infty }{{{e}^{-i\left( {{\omega }_{1}}+{{\omega }_{2}}-\omega  \right)t}}dt}=\delta \left( {{\omega }_{1}}+{{\omega }_{2}}-\omega  \right)\lb{20}
\eeq
and find
\beq
{{\left( \hat{a}\delta {{{\hat{N}}}_{e}} \right)}_{\omega }}=\int\limits_{-\infty }^{\infty }{\hat{a}({{\omega }_{1}})\delta {{{\hat{N}}}_{e}}}({{\omega }_{2}})\delta \left( {{\omega }_{1}}+{{\omega }_{2}}-\omega  \right)\frac{d\omega_1d\omega_2}{\left(2\pi\right)^{1/2}}.        \lb{21}
\eeq
Now we take the integral over $d\omega_2$ in Eq.~\rf{21} and come to
\beq
{{\left( \hat{a}\delta {{{\hat{N}}}_{e}} \right)}_{\omega }}=\frac{1}{{{\left( 2\pi  \right)}^{1/2}}}\int\limits_{-\infty }^{\infty }{\hat{a}({{\omega }_{1}})\delta {{{\hat{N}}}_{e}}}(\omega -{{\omega }_{1}})d{{\omega }_{1}}.\lb{22}
\eeq
Therefore $\left( \hat{a}\delta\hat{N}_e \right)_{\omega }$ is a convolution of operators $\hat{a}(\omega )$ and $\delta \hat{N}_e(\omega)$. Similar way we find
\beq
{{\left( \hat{a}^+\delta {{{\hat{N}}}_{e}} \right)}_{\omega }}=\frac{1}{{{\left( 2\pi  \right)}^{1/2}}}\int\limits_{-\infty }^{\infty }{\hat{a}^+({-{\omega }_{1}})\delta {{{\hat{N}}}_{e}}}(\omega -{{\omega }_{1}})d{{\omega }_{1}}.\lb{23}
\eeq
Now we express the mean $M=\left\langle\hat{a}^+(t)\delta \hat{N}_e(t)\hat{a}(t)\delta\hat{N}_e(t)\right\rangle$ through  Fourier-components of $\hat{a}^+(t)$, $\hat{a}(t)$ and $\delta \hat{N}_e(t)$. First, we write
\beq
M=\frac{1}{2\pi }\int\limits_{-\infty }^{\infty }{\left\langle {{\left( {{{\hat{a}}}^{+}}\delta {{{\hat{N}}}_{e}} \right)}_{{{\omega }_{1}}}}{{\left( \hat{a}\delta {{{\hat{N}}}_{e}} \right)}_{{{\omega }_{2}}}} \right\rangle {{e}^{-i\left( {{\omega }_{1}}+{{\omega }_{2}} \right)t}}d{{\omega }_{1}}d{{\omega }_{2}}}.\lb{24}
\eeq
We insert Eqs.\rf{22} and \rf{23} into Eq.\rf{24} and obtain
\begin{widetext}\beq
M=\frac{1}{{{\left( 2\pi  \right)}^{2}}}\int\limits_{-\infty }^{\infty }{\left\langle \int\limits_{-\infty }^{\infty }{{{{\hat{a}}}^{+}}(-{{\omega }_{1}}')\delta {{{\hat{N}}}_{e}}}({{\omega }_{1}}-{{\omega }_{1}}')d{{\omega }_{1}}'\int\limits_{-\infty }^{\infty }{\hat{a}({{\omega }_{1}}'')\delta {{{\hat{N}}}_{e}}}({{\omega }_{2}}-{{\omega }_{1}}'')d{{\omega }_{1}}'' \right\rangle {{e}^{-i\left( {{\omega }_{1}}+{{\omega }_{2}} \right)t}}d{{\omega }_{1}}}d{{\omega }_{2}}.\lb{25}
\eeq\end{widetext}
The laser at low excitation does not generate coherent radiation, $\left<\hat a\right> = 0$, $\left<\hat{v}\right> = 0$, so it follows from Eq.~\rf{MBE_St2} that $\left\langle \hat{a}(t)\delta {{{\hat{N}}}_{e}}(t) \right\rangle =0$. Then applying  the cumulant-neglect closure method \ct{Wu_1984,10.1115/1.3173083} in Eq.~\rf{25} we write
\[
  \left\langle {{{\hat{a}}}^{+}}(-{{\omega }_{1}}')\delta {{{\hat{N}}}_{e}}({{\omega }_{1}}-{{\omega }_{1}}')\hat{a}({{\omega }_{1}}'')\delta {{{\hat{N}}}_{e}}({{\omega }_{2}}-{{\omega }_{1}}'') \right\rangle \approx\]\beq \hspace{-0.4cm}\left\langle {{{\hat{a}}}^{+}}(-{{\omega }_{1}}')\hat{a}({{\omega }_{1}}'') \right\rangle \left\langle \delta {{{\hat{N}}}_{e}}({{\omega }_{1}}-{{\omega }_{1}}')\delta {{{\hat{N}}}_{e}}({{\omega }_{2}}-{{\omega }_{1}}'') \right\rangle,  \lb{26} 
\eeq
taking into account that operators $\hat{a}$ and $\hat{a}^+$ commute with $\delta\hat{N}_e$.
Relation \rf{26} reminds the cluster expansion for correlations in the time domain \ct{PhysRevA.75.013803} when 
\beq
\left<\hat{a}^+\hat{a}\delta \hat{N}_e^2\right> \approx \left<\hat{a}^+\hat{a}\right>\left<\delta \hat{N}_e^2\right> + 2\left<\hat{a}^+\delta \hat{N}_e\right>\left<\hat{a}\delta \hat{N}_e\right>. \lb{CLM_0}
\eeq
For the laser with a low excitation the second term on the right in Eq.~\rf{CLM_0} is zero so 
\beq
\left<\hat{a}^+\hat{a}\delta \hat{N}_e^2\right> = \left<\hat{a}^+\hat{a}\right>\left<\delta \hat{N}_e^2\right>. \lb{CLM_1}
\eeq
Eq.~\rf{26} is a "cluster expansion"  for Fourier component operators. 

According with Eqs.~\rf{13} and \rf{15}
\beqr
  & \left\langle {{{\hat{a}}}^{+}}(-{{\omega }_{1}}')\hat{a}({{\omega }_{1}}'') \right\rangle =n({{\omega }_{1}}')\delta ({{\omega }_{1}}'+{{\omega }_{1}}''), \nonumber\\ 
 & \left\langle \delta {{{\hat{N}}}_{e}}({{\omega }_{1}}-{{\omega }_{1}}')\delta {{{\hat{N}}}_{e}}({{\omega }_{2}}-{{\omega }_{1}}'') \right\rangle =&\lb{27}\\&\delta^2N_e({{\omega }_{1}}-{{\omega }_{1}}')\delta ({{\omega }_{1}}-{{\omega }_{1}}'+{{\omega }_{2}}-{{\omega }_{1}}''). \nonumber
\eeqr
We insert Eq.~\rf{27} into Eq.~\rf{26}; Eq.~\rf{26} into Eq.~\rf{25}, carry out the integration in Eq.~\rf{25} taking into account delta-functions and come to 
\[
M=\frac{1}{\left( 2\pi  \right)^2}\int\limits_{-\infty}^{\infty}\left( \int\limits_{-\infty }^{\infty }n(\omega ')\delta^2N_e(\omega_1-\omega_1')d\omega ' \right)d\omega \]
\beq
=\frac{1}{2\pi }\int_{-\infty }^{\infty }S_{aN_e}(\omega )d\omega . \lb{Conv_int}
\eeq
We see from Eq.~\rf{Conv_int} that the spectrum ${{S}_{a{{N}_{e}}}}(\omega )$ of the operator product $\hat{a}(t)\delta {{\hat{N}}_{e}}(t)$ 
\beq
S_{aN_e}(\omega )=\frac{1}{2\pi }\int_{-\infty }^{\infty }n(\omega ')\delta^2N_e(\omega_1-\omega_1')d\omega' 
\lb{conv_0}
\eeq
is a convolution of spectra $n(\omega)$ and $\delta^2N_e(\omega)$ of operators $\hat{a}(t)$ and $\delta \hat{N}_e(t)$.

The structure of  formula \rf{conv_0}  and the interpretation of $n(\omega)$ as a probability density (see Eq.~\rf{prob1}) points out on the  interpretation of ${{S}_{a{{N}_{e}}}}(\omega )$. We calculate ${\bar{S}_{a{{N}_{e}}}}=(2\pi)^{-1}\int_{-\infty }^{\infty }{{S}_{a{{N}_{e}}}}(\omega )d\omega$ and, by the analogy with Eq.~\rf{prob1}, define the probability
\beq
        dp_{aN_e}(\omega) = S_{aN_e}(\omega)d\omega/(2\pi\bar{S}_{aN_e}). \lb{prob_conv}
\eeq
This is the probability of the event, that an emitter and the field are in the band of states with the total energy of the emitter and the field in the interval from $\hbar(\omega_0+\omega)$ to $\hbar(\omega_0+\omega+d\omega)$, and $S_{aN_e}(\omega)/(2\pi\bar{S}_{aN_e})$ is the probability density for such event. 

Now we will comment our perturbation approach. In order to find some mean value, as the mean photon number $n$, we do not need to solve time-dependent equations \rf{MBE_0} for operators. It is enough to calculate the spectrum $n(\omega)$  and use Eq.~\rf{14}.  So instead of the linearization of equations of motion for operators, we approximately calculate spectra with the help of Eq.~\rf{conv_0}. We calculate the field spectrum $n(\omega)$  neglecting by the population fluctuations, which is a zero-order approximation in the perturbation approach. The spectrum $\delta^2N_e(\omega)$ of the population fluctuations will be found using results of the the zero-order approximation. Then, when we know $n(\omega)$ and $\delta^2N_e(\omega)$ (though approximately), we will use Eq.~\rf{conv_0} for calculations of the spectrum $S_{aN_e}(\omega)$ of the operator product $\hat{a}(t)\delta {{\hat{N}}_{e}}(t)$. Knowing $S_{aN_e}(\omega)$  we can find from Eqs.~\rf{FC_1} and \rf{FC_2} any spectrum  and  mean value in the first order on population fluctuations and in the stationary case. The procedure may be repeated in the higher-order approximations.  

In order to preserve commutation relations for Bose-operators of the lasing mode we  calculate  corrections to zero-order diffusion coefficients. 
\section{\label{DC_app}Diffusion coefficients}
Generalized Einstein relations \ct{Lax_book1966} for the polarization of emitters lead to
	\[\left\langle \frac{d}{dt}{{{\hat{v}}}^{+}}\hat{v} \right\rangle =-{{\gamma }_{\bot }}\left\langle {{{\hat{v}}}^{+}}\hat{v} \right\rangle +2{{D}_{{{v}^{+}}v}}=\]\vspace{-0.75cm}\beq f\left\langle \frac{d}{dt}{{{\hat{N}}}_{e}} \right\rangle =f{{\gamma }_{\parallel }}\left( P{{N}_{g}}-{{N}_{e}} \right) \lb{Cen_En_rel}\eeq
so the diffusion coefficient
\beq
	2{{D}_{{{v}^{+}}v}}=f\left[ {{\gamma }_{\bot }}{{N}_{e}}+{{\gamma }_{\parallel }}\left( P{{N}_{g}}-{{N}_{e}} \right) \right]. 	\lb{Dif_c_ex1}\eeq
Similar way we find
	\beq2{{D}_{v{{v}^{+}}}}=f\left[ {{\gamma }_{\bot }}{{N}_{g}}-{{\gamma }_{\parallel }}\left( P{{N}_{g}}-{{N}_{e}} \right) \right].\lb{Dif_c_ex2}\eeq 	
Using the energy conservation law \rf{eql_1} we write
	\[
	2{{D}_{{{v}^{+}}v}}=f{{\gamma }_{\bot }}\left[ {{N}_{e}}+\left( 2\kappa /{{\gamma }_{\bot }} \right)n \right],\]\vspace{-0.75cm}\beq   2{{D}_{v{{v}^{+}}}}=f{{\gamma }_{\bot }}\left[ {{N}_{g}}-\left( 2\kappa /{{\gamma }_{\bot }} \right)n \right]. \lb{Dif_c_correct}\eeq
Using diffusion coefficients \rf{Dif_c_correct}  we calculate
\beq
	\left\langle \left[ {{{\hat{a}}}_{0}},\hat{a}_{0}^{+} \right] \right\rangle =1+\frac{\left( 4\kappa /{{\gamma }_{\bot }} \right)n}{\left( 1+2\kappa /{{\gamma }_{\bot }} \right)\left( {{N}_{th}}-N \right)}\lb{dif_c_brake}\eeq 	
So diffusion coefficients \rf{Dif_c_correct} break Bose commutation relations for  ${{\hat{a}}_{0}}$ and they cannot be used in the zero-order approximation and we must use $2{{D}^{(1)}_{{{v}^{+}}v}}$ and $2{{D}^{(1)}_{{{v}}v^{+}}}$ given by Eq.~\rf{d_coeff_1} with $N_1$ given by Eq.~\rf{N1_expr}.

Without population fluctuations, when $\left\langle \frac{d}{dt}{{{\hat{N}}}_{e}} \right\rangle =0$ in Eq.~\rf{Cen_En_rel}, we have $2D^{(0)}_{{{v}^{+}}v}=f{{\gamma }_{\bot }}{{N}_{e}}$ and $2D^{(0)}_{v{{v}^{+}}}=f{{\gamma }_{\bot }}{{N}_{g}}$. It is shown in the main text that such zero-order diffusion coefficients  preserve commutation relations $\left\langle \left[ {{{\hat{a}}}_{0}},\hat{a}_{0}^{+} \right] \right\rangle =1$. 

The sum of diffusion coefficients \rf{Dif_c_correct}
\beq
2{{D}_{{{v}^{+}}v}} + 2{{D}_{vv^+}} = f\gamma_{\perp}N_0 \lb{sum_dif_coef}
\eeq
does not depend on the population fluctuations, the same must be true for the sum $2{{D}^{(1)}_{{{v}^{+}}v}} + 2{{D}^{(1)}_{vv^+}}$, this is why we chose the same $N_1$ in diffusion coefficients \rf{d_coeff_1}. 
\section{\label{Sigma_eqs}Equations for population fluctuations.}
Using Eqs.~\rf{MBE_0} and the usual rule of the differentiation of products we write equations for   $\hat{\Sigma}$, given by Eq.~\rf{sigma_def}, $\hat{n}=\hat{a}^+\hat{a}$ and
$\hat{D}=f^{-1}\sum_{i\neq j}\hat{v}_i^+\hat{v}_i$. Neglecting   population fluctuations we replace  population operators $\hat{N}_{e,g}$ by their means  ${N}_{e,g}$ and obtain  
\begin{subequations}\lb{binary_eqs0}\beqr
  \dot{\hat{n}} &=&-2\kappa \hat{n}+\Omega \hat{\Sigma }+{{{\hat{F}}}_{n}} \lb{binary_eqs0_1}\\ 
 \dot{\hat{\Sigma }}&=&-\left( \kappa +{{\gamma }_{\bot }}/2 \right)\hat{\Sigma }+ \lb{binary_eqs0_2}\\
& & \hspace{1cm}2\Omega f \left( \hat{n}N+\hat{D}+N_e \right)+{{{\hat{F}}}_{\Sigma }}\nonumber\\
 \dot{\hat{D}}&=&-{{\gamma }_{\bot }}\hat{D}+\Omega N\hat{\Sigma }+\hat{F}_D, \lb{binary_eqs0_3} 
\eeqr\end{subequations}
where $N=N_e-N_g$. Non-zero diffusion coefficients $2D_{\alpha\beta}$, $\alpha,\beta = \{n,\Sigma,D\}$ in correlations of Langevin forces 
$
\left<\hat{F}_{\alpha}(t)\hat{F}_{\beta}(t')\right> = 2D_{\alpha\beta}\delta{(t-t')}
$
are
\beqr
&2{{D}_{nn}}=2\kappa n, \hspace{0.5cm} 2{{D}_{\Sigma \Sigma }}=f[2\kappa D+{{\gamma }_{\bot }}N_0n+\left( 2\kappa +{{\gamma }_{\bot }} \right)N_e]&\nonumber\\
&2{{D}_{DD}}={{\gamma }_{\bot }}(N_0D+2N_eN_g),&\lb{dif_n0}\\
&2D_{\Sigma n} = 2D_{n\Sigma} = \kappa\Sigma, \hspace{0.5cm} 2D_{\Sigma D} = 2D_{D\Sigma} = (\gamma_{\perp}/2)N_0\Sigma.\nonumber&
\eeqr
Diffusion coefficients \rf{dif_n0} are the same as ones found from the generalized Einstein relations \ct{Lax_book1966}, apart of the  term $\sim 2N_eN_g$ in $2{{D}_{DD}}$, this term must be added when we neglect population fluctuations.  The derivation of  diffusion coefficients \rf{dif_n0} will be presented in the forthcoming paper.  

We separate mean values and fluctuation operators in $\hat{n}$, $\hat{\Sigma}$ and $\hat{D}$
\beq
        \hat{n} = n+\delta \hat{n}, \hspace{0.5cm}
        \hat{\Sigma} = \Sigma+\delta \hat{\Sigma}, \hspace{0.5cm} \hat{D} = D+\delta\hat{D}, \lb{fl_means}
\eeq
insert \rf{fl_means} into Eqs.~\rf{binary_eqs0} and obtain equations for mean values  
\begin{subequations}\lb{binary_eqs_st}\beqr
  0 &=&-2\kappa n+\Omega \Sigma \lb{binary_eqs_st_1}\\ 
 0&=&-\left( \kappa +{{\gamma }_{\bot }}/2 \right)\Sigma+2\Omega f \left( nN+D+N_e \right) \lb{binary_eqs_st_2}\\
 0&=&-{{\gamma }_{\bot }}D+\Omega N\Sigma \lb{binary_eqs_st_3} 
\eeqr\end{subequations}
and for fluctuation operators $\delta\hat{n}$, $\delta\hat{\Sigma }$ and $\delta\hat{D}$
\begin{subequations}\lb{binary_eqs_fl}\beqr
  \delta\dot{\hat{n}} &=&-2\kappa \delta\hat{n}+\Omega \delta\hat{\Sigma }+{{{\hat{F}}}_{n}} \lb{binary_eqs_fl_1}\\ 
 \delta\dot{\hat{\Sigma }}&=&-\left( \kappa +{{\gamma }_{\bot }}/2 \right)\delta\hat{\Sigma }+\lb{binary_eqs_fl_2}\\& & \hspace{1cm}2\Omega f \left( \delta\hat{n}N+\delta\hat{D} \right)+{{{\hat{F}}}_{\Sigma }} \nonumber\\
 \delta\dot{\hat{D}}&=&-{{\gamma }_{\bot }}\delta\hat{D}+\Omega N\delta\hat{\Sigma }+\hat{F}_D. \lb{binary_eqs_fl_3} 
\eeqr\end{subequations}
Solving linear Eqs.~\rf{binary_eqs_fl} by Fourier-transform we obtain $\delta\hat{\Sigma}(\omega)$. With  $\delta\hat{\Sigma}(\omega)$ and diffusion coefficients \rf{dif_n0} we find the spectrum
$\delta^2{\Sigma}(\omega)$
\beq
\left<\delta\hat{\Sigma}(\omega)\delta\hat{\Sigma}(\omega')\right> = \delta^2{\Sigma}(\omega)\delta(\omega+\omega').
\lb{Sigma_spect} 
\eeq
 
\bibliography{myrefs}

\end{document}